\documentclass[
 preprint,
showpacs,preprintnumbers,
showkeys,
nofootinbib,
 amsmath,amssymb,
 aps,
prd,
 longbibliography,
 lengthcheck
]{revtex4-1}
\usepackage{fleqn}
\usepackage{epsfig}
\usepackage{graphicx}
\usepackage{dcolumn}
\usepackage{bm}
\usepackage{longtable}
\usepackage{lscape}

\def\beq{\begin{equation}}
\def\be{\begin{equation}}
\def\eeq{\end{equation}}
\def\ee{\end{equation}}
\def\bea{\begin{eqnarray}}
\def\eea{\end{eqnarray}}
\def\bq{\begin{quote}}
\def\eq{\end{quote}}

\def\beqn{\begin{eqnarray}}
\def\eeqn{\end{eqnarray}}

\begin{document}

\title{Hints for the existence of hexaquark states in the baryon-antibaryon sector}
\author{Mario Abud and Franco Buccella}
\affiliation{Dipartimento di Scienze Fisiche, Universit\`{a} di Napoli
``Federico II" and INFN, Sezione di Napoli, Via Cintia, I-80126 Napoli,
Italy}

\author{Francesco Tramontano}
\affiliation{Theory Group, Physics Department,
CERN CH-1211, Geneva 23,
Switzerland}

\begin{abstract}
The discovery of some baryon-antibaryon resonances has led us to consider 
$3q~3\bar{q}$ systems as possible candidates. We predict their
spectrum in the framework of a constituent model, where the
chromo-magnetic interaction plays the main role. The relevant
parameters are fixed by the present knowledge on tetraquarks. The emerging
scenario complies well with experiment, besides the description of the baryon-antibaryon
resonances, we find evidence for new tetraquark states, namely the $a_{0}(Y)$ in the hidden
strangeness sector and, in the $cs\overline{c}\overline{s}$ sector, the $Y(4140)$ and the
$X(4350)$. A detailed account of the spectra and the decay channels is provided for future
comparisons with data. 
\end{abstract}

\preprint{DSF-NA 15-2009}
\preprint{CERN-PH-TH/2009-246}



\maketitle

\section{Introduction}

The presence in the hadron spectrum of mesons consisting of two~$q$'s and
two~$\bar{q}$'s~\cite{JafPHEN1,J,JafPHEN2} as well as of baryons consisting of~$4q$
and a~$\bar{q}$ has been considered since many years~\cite{HS}.

A long time ago Jaffe proposed that the lightest scalar states,
$f_{0}/\sigma$, $\kappa$ together with the rest of their nonet,
should be interpreted as~$qq\bar{q}\bar{q}$ states~\cite{JafPHEN1}.

The simplifying assumption~\cite{JW} of considering only~$2q$ pairs
transforming as a ($\bar{3}_{c},1_{s},\bar{3}_{F}$) representation of
$SU(3)_{c}\times SU(2)_{s}\times SU(3)_{F}$, straitens the whole spectrum to
the lightest scalar nonet, namely $f_{0}(600)$, $\kappa(800)$ and
$f_{0}/a_{0}(980)$ as built with a pair of such a diquark and
anti-diquark~\cite{MPPR}. This interpretation was recently enforced by
experiments confirming the presence of hidden strangeness in both the states
$f_{0}(980)$ and $a_{0}(980)$~\cite{KLOE}, promoting the tetraquarks to a more
solid status.

Candidates with open or hidden charm come from the study of non-leptonic B decays
at BABAR and BELLE, as anticipated in~\cite{BigiMaiani}, and from BES.

In this paper we study the spectrum of the states consisting of three quarks
and three antiquarks in S-wave, interacting via chromo-magnetism.
Besides strangeness we include also charm and assume for chromo-magnetism 
its full content~\cite{DGG}, treated along the lines of ref.~\cite{BHRS}. 
  
It happens that this hypothesis can successfully interpret some observed
baryon-antibaryon negative parity states in~$p\overline{p}$~\cite{Bai06},
$\Lambda_{c}\overline{p}$~\cite{Gaby}
and $\Lambda_{c}\overline{\Lambda}_{c}$~\cite{Y4660}, assuming for the
parameters (constituent masses and effective couplings) those values
obtained from the tetraquarks phenomenology. To study the case of broken
flavor symmetry we had to resort to machine computation.

The paper is organized as follows: in section~I we introduce the basics of
chromo-magnetism with a formulation more suitable for algebraic computation.
Section~II deals with the formalism for the construction of the tetraquarks
states and the study of the open door decays. In section~III, IV and V we
discuss the phenomenology of tetraquarks states and the parameter fixing 
of the model. Hexaquark states are introduced in sections~VI along with the
details entering the calculation. In section~VII we present the results we
found for the spectrum and compare them with the relevant experimental data.
Section~VIII contains our conclusions.
Finally, the appendix~A contains a table with the full spectrum of the
baryon-antibaryon systems that were taken under consideration, while in
appendix~B the crossing matrices required for the study of the decays
of tetraquarks are reported. The matrix elements of the chromo-magnetic
operator are given, for all cases, in appendix~C.

\section{The chromo-magnetic interaction}

The hyperfine interaction arising from one gluon exchange between constituents
leads to a simple Hamiltonian involving the color and spin degrees of freedom:
\begin{equation}
H_{CM}=\sum_{i}\,m_{i}-\sum_{i<j}\,C_{ij}\,O_{CM}^{(i,\ j)}\label{eq:HCM}
\end{equation}
the index $i~(j)$ refers to the $i$th ($j$th) quark, $m_{i}$ its mass and
$C_{ij}$ appropriate coupling constants. The kinetic energy is absorbed in the
mass term, so it is not surprise that the quarks masses depend on the system
under consideration. The $C_{ij}$'s depend not only on the $m_{i}$'s (as
$1/m_{i}m_{j}$) but also on the wave function at zero distance of the pair
$(i,\ j)$, so depending on the system as well. Chromo-Magnetism (CM) is encoded in
$O_{CM}^{(i,\ j)}$, the two particles chromo-magnetic operator, which is given
by:
\begin{equation}
O_{CM}^{(i,\ j)}=\frac{1}{4}\,\sum_{a=1}^{8}\,\sum_{k=1}^{3}\,(\lambda
_{a}\,\otimes\,\sigma_{k})^{(i)}\,(\lambda_{a}\,\otimes\,\sigma_{k})^{(j)}
\label{eq:Oij}
\end{equation}
where $\lambda_{a}$ are the Gell-Mann matrices and $\sigma_{k}$ the Pauli matrices.
It is reminiscent of the well known exchange interaction and can be expressed in terms
of permutation operators for color and spin $P_{c}^{(i,j)},\ P_{s}^{(i,j)}$ respectively.
The action on a $(i,j)$ quark-quark (antiquark-antiquark) pair is given by
\begin{equation}
O_{CM}^{qq}=(P_{c}-1/3)\otimes\,(P_{s}-1/2) \label{eq:Permt}
\end{equation}
where $P_{c}^{(i,j)}$ and $P_{s}^{(i,j)}$ exchange the colors and spins
(acting independently), of the pair ${(i,j)}$.
Eigenvectors of~\ref{eq:Permt} are the diquark states of definite symmetry
in color and spin
$(6,3)(SS),~(6,1)(SA),~(\overline{3},3)(AS),~(\overline{3},1)(AA)$
with eigenvalues $(-1/3,1,2/3,-2)$ respectively.

To express the result for a quark-antiquark pair it is useful to define
a generic $T_{N}$ for the group $SU(N)$ as the object:
$T_{N}:\Psi _{A}\Xi ^{B}\rightarrow 1/N\ \Psi _{A}\Xi ^{B}-\delta _{A}^{B}
\Psi _{C}\Xi ^{C}$, with $\Psi _{A}$ in the representation $N$
and $\Xi ^{B}$ in the c.c. representation $\overline{N}$.
Making the identification $N=3$ for $T_{c}\ $and$\ N=2\ $ for $T_{s}$
we can write quite simply:%
\begin{equation}
O_{CM}^{q\overline{q}}=-T_{c}\otimes\,T_{s}\label{eq:TT}%
\end{equation}
The eigenvectors of $T_{N}$ are the singlet representation
($\delta_{A}^{B}\Psi_{c}\Xi^{C}$) with eigenvalue ($1/N-N$) and the
adjoint representation ($\Psi_{A}\Xi^{B}-1/N\delta_{A}^{B}\Psi_{c}\Xi^{C}$)
with eigenvalue $1/N$. So eigenvectors and eigenvalues of the chromo-magnetic
operator in the present case are: $(8,3),(8,1),(1,3),(1,1)$ with eigenvalues
$(-1/6,1/2,4/3,-4)$ respectively.

By far the more bonded diquark is the $(\overline{3},1)(AA)$ whose $SU(3)_{F}$
flavor content, as dictated by the Pauli principle, is $\overline{3}_{F}$.
This is the so called good diquark, it transforms as a scalar antiquark.
If one assumes the hypothesis of Jaffe and Wilczek~\cite{JW}, the
spectrum of the tetraquarks remains restricted to the scalar nonet suggested by
Jaffe a long time ago. The vector, or bad diquark $(\overline{3},3)(AS)$, allows
for higher spin states but, since it is a $6_{F},$ it also introduces exotics,
i.e. multiplets higher than $SU(3)_{F}$ nonets and are excluded from most
models. The other two $6_{c}$ states, that Jaffe~\cite{JafPHEN2,JafPR}
called sometimes ``worse'' are not in general taken into account neither.

In the present approach in searching for the eigenstates of the
chromo-magnetic operator we do not truncate the space in any way, such that,
in some sense, all four possible diquarks enter the game. 

It is easy to see that we have the following spin-flavor multiplets:
spin~0 has four nonets and two $27_{F}$'s ,
spin~1 has two nonets, four octets, one $27_{F}$, two decuplets and two
antidecuplets, finally spin~2 has two nonets and one $27_{F}$. 
Exotics, as $I=2$ states, are not excluded a priori but we
think that these states are much less stable and difficult to be observed.

Often, we have found a number of near threshold decays, usually attributed as
molecular states, that are well described by chromo-magnetism. In particular
the introduction of the $(6,3)$ diquark encompass the dichotomy between
diquark and molecular models as clearly argued in~\cite{Sch08}.

They showed that the molecular state is not an independent state, but is a
linear combination of $(\overline{3},1)(3,1)$ and $(6,3)(\overline{6},3)$, the later $(6,3)$,
 by the way, is the only other diquark with negative chromo-magnetic energy ($-1/3$). 
Their observation indicates that a minimal
diquark model should include both pairs, and interestingly enough, it would
comprise all spin cases as S-wave tetraquarks lying in only $SU(3)_{F}$
nonets. From the point of view of $SU(6)_{cs}$ this means that a diquark
should transform as the symmetric representation, $21$ (so as
$\overline{3}_{F}$).

A purely phenomenological motivation to include the $(6,3)$ diquark 
is that the mass of the $\bar{3},S=0,(ud)_{I=0}$ pair, say $\mu$,
is related to the mass of the $\Lambda$ hyperon by the
relation~\footnote{We are indebted to Prof. P. Minkowski for bringing
this remark to our knowledge}: $\mu = m_{\Lambda} - m_s$,
which for a state consisting of two of these objects which have no
mutual chromo-magnetic interaction imply about twice the mass of the
$f_{0}(600)$.
Instead, by considering the vector space consisting of both the
$(\bar{3},1) (3, 1)$ and $(6, 3) (\bar{6}, 3),~S = 0$
color singlet states, the lightest state has a binding energy about $2.7$
times larger than the diagonal matrix element for
$(\bar{3},1) (3,1)$~\cite{BHRS}.

In the flavor symmetry limit, i.e. when the couplings $C_ {ij}$ are all
equal to each other, it is well known that $O_{CM}$ can be expressed as
a combination of Casimirs. This fact has been extensively exploited in the
pioneering works of Jaffe~\cite{JafPHEN2} and in many other
works~\cite{HS}.
In the present paper we shall attack the more complicated issue of considering
different masses and couplings, in most of such cases we have to rely on
symbolic manipulations that we performed with FORM~\cite{form}.
The expressions in Eqs.~(\ref{eq:Permt}) and (\ref{eq:TT}) result quite
suitable for computer implementation.
 
\section{``Open door'' channels for tetraquarks}

It has been observed for the first time by Jaffe~\cite{JafPHEN1} that
$qq\bar{q}\bar{q}$ mesons may decay into two ordinary (i.e. color singlet)
mesons PP, PV, VV (P stands for a pseudoscalar and V for a vector)
by simply separating from each other, as long as it is kinematically
allowed. He called these channels ``open door'' or ``Ozi super-allowed''
decays, since they can occur without gluon exchange or quark annihilation.
In open door channels, S-wave states have to decay into S-wave mesons with
zero relative angular momentum.

In general calculations are performed in the diquark-antidiquark basis, i.e.
the tetraquark is represented as $q_{1}q_{2}\overline{q}_{3}\overline{q}_{4}$
denoted~\textbf{[12,34]} in the following. Evidently the diquark and the
antidiquark cannot separate from each other as they can never be color
singlets. So, in order to access the open door channels it is convenient
to pass to the meson-meson basis \textbf{[13,24]} and \textbf{[14,23]}
which, obviously, coincide if antiquarks 3 and 4 have the same flavor.

In order to have some uniformity in the conventions, we maintain those
of~\cite{BHRS}. We call the basis for spin~0: as $\phi$ in \textbf{[12,34]},
$\alpha$ in \textbf{[13,24]} and $\epsilon$ in \textbf{[14,23]},
in the same order one has $\psi,~\beta$ and $\chi$ for spin 1,
while those of spin 2 are called $\xi,~\gamma$ and $\delta$.
To characterize each basis, we have only to specify the color-spin content of
the first and second pairs in the brackets, which combine to form the color
singlets i.e. the set of physical states.
\begin{center}
\textbf{Spin 0}
\end{center}
\begin{center}
\begin{tabular}{rll}
$(\phi)\textbf{[12,34]}:     $&$ [(6,3)(\overline{6},3)]; $&$ [(\overline{3},1)(3,1)]; $ \\
                              &$ [(6,1)(\overline{6},1)]; $&$ [(\overline{3},3)(3,3)] $ \\
$ $ \\
$(\alpha)\textbf{[13,24]}:   $&$ [(1,1)(1,1)];            $&$  [(1,3)(1,3)]; $ \\
                              &$ [(8,1)(8,1)];            $&$ [(8,3)(8,3)]$ \\
$ $ \\
$(\epsilon)\textbf{[14,23]}: $& as $ \textbf{[13,24]}  $& \\
\end{tabular}
\begin{equation} \label{eqs0:4}
\end{equation}
\end{center}
For $\alpha$ and $\epsilon$ the first components are $PP$ and the second $VV$.
The last two are $P_{8}P_{8}$ and $V_{8}V_{8,}$ where $P_{8}$ is a
colored pseudoscalar and $V_{8}$ a colored vector
\begin{center}
\textbf{Spin 1}
\end{center}
\begin{center}
\begin{tabular}{rlll}
$(\psi)\textbf{[12,34]}:     $&$ [(6,3)(\overline{6},3)]; $&$ [(\overline{3},3)(3,3)]; $&$ [(\overline{3},1)(3,3)]; $\\
                              &$ [(6,3)(\overline{6},1)]; $&$ [(\overline{3},3)(3,1)]; $&$ [(6,1)(\overline{6},3)]  $\\

$ $ \\
$(\beta)\textbf{[13,24]}:    $&$ [(1,1)(1,3)];            $&$ [(1,3)(1,1)];            $&$ [(1,3)(1,3)]; $ \\
$ $ \\ 
                             &$ [(8,1)(8,3)];            $&$ [(8,3)(8,1)];            $&$ [(8,3)(8,3)] $ \\
$ $ \\
$(\chi)\textbf{[14,23]}:     $& as $ \textbf{[13,24]}  $& \\
\end{tabular}
\begin{equation} \label{eqs1:4}
\end{equation}
\end{center}
So $\beta_{1}$, $\chi_{1}$ ($\beta_{2}\ \chi_{2}$) are PV(VP) and $\beta_{3}$, $\chi_{3}$ are VV.
\begin{center}
\textbf{Spin 2}
\end{center}
\begin{center}
\begin{tabular}{rll}
$(\xi)\textbf{[12,34]}:     $&$ [(6,3)(\overline{6},3)];  $&$ [(\overline{3},3)(3,3)] $ \\
$ $ \\
$(\gamma)\textbf{[13,24]}:  $&$ [(1,3)(1,3)];             $&$ [(8,3)(8,3)]            $ \\
$ $ \\
$(\delta)\textbf{[14,23]}:  $& as $ \textbf{[13,24]}  $& \\
\end{tabular}
\begin{equation} \label{eqs2:4}
\end{equation}
\end{center}
The only open door channel for a tensor meson is, evidently, VV. 

The relative probability for the particle decaying through a specific channel
is given by the square of the corresponding component of the normalized
eigenvector of the state multiplied by phase space (as is assumed all dynamical
amplitudes to be the same). For convenience we call the square of the component
along the channel the probability factor (PF) for that channel.
In some cases we have also to consider the non open door channels,
if for instance, the open door have negligible probabilities or are
kinematically forbidden, and so violations of the OZI rule would enter
the game. In particular the $P_{8}P_{8}$ or $V_{8}V_{8}$ channel can become
relevant at order $O(\alpha_{s})$, as the exchange of one gluon in the
t-channel converts this object into an ordinary $PP$ or $VV$ pairs.

The so called crossing matrices operating the change of a basis into another,
arise from well known Fierz identities for color and spin~\cite{JafPHEN2} and
are available in many places, for definiteness we will refer to~\cite{BHRS}.
They are reproduced, together with a necessary completion, in Eqs.~[\ref{cros01}-\ref{cros2}].

\section{Tetraquark States}

It is immediate to realize that the overall chromo-magnetic contribution in
Eq.~\ref{eq:HCM} (let us call it $O_{CM}$ and assume thoroughly 
$C_{q\overline{q^{\prime}}}=C_{qq^{\prime}}$ for any (anti) quarks pair)
greatly simplifies for $0^{+}$ and $2^{+}$ states made of at least three constituents
with the same flavor, say of type $q\overline{q}q\overline{q}^{\prime}$ ($q$ is not
necessarily a light quark and $q$ and $q\prime$ can incidentally coincide), since
the corresponding matrices depend exclusively on the combination
$(C_{qq}+C_{qq^{\prime}})$, which factorizes out. For $2^{+}$ we have:
$O_{CM}=-4/3(C_{qq}+C_{qq^{\prime}})\,diag(1,1)$, while for $0^{+}$:
\begin{eqnarray}
O_{CM}&=&-1/2(C_{qq}+C_{qq^{\prime}}) \cdot \\
&& \left(
\begin{array}[c]{llll}
8 & 0 & 0 & -4\sqrt{\frac{2}{3}}\\
0 & -\frac{8}{3} & -4\sqrt{\frac{2}{3}} & 0\\
0 & -4\sqrt{\frac{2}{3}} & -1 & -\frac{5}{\sqrt{3}}\\
-4\sqrt{\frac{2}{3}} & 0 & -\frac{5}{\sqrt{3}} & \frac{19}{3}
\end{array} \nonumber
\right). \label{M0}
\end{eqnarray}
The eigenvalues of the above matrix are $\lambda_{1}=1/3(17+\sqrt{241})$,
$\lambda_{2}=1/3(\sqrt{241}-1),~\lambda_{3}=1/3(17-\sqrt{241}),~\lambda
_{4}=-1/3(\sqrt{241}+1)$, with corresponding eigenvectors (for briefness we
give decimal approximations) $(-0.74,0.04,0.17,0.65)$, $(0.64,0.18,-0.41,0.62)$,
$(0.18,-0.64,0.62,0.41)$ and $(0.04,0.74,0.64,0.17)$.

The spectrum is given by
$M_{a}^{(0)}=3m_{q}+m_{q^{\prime}}-1/2\ \lambda_{a}\ (C_{qq}+C_{qq^{\prime}})$,
$(a=1,...,4)$ for $0^{+}$ and by
$M_{b}^{(2)}=3m_{q}+m_{q^{\prime}}+4/3\ (C_{qq}+C_{qq^{\prime}}),(b=1,2)$ for
$2^{+}$.
These considerations apply also to the case of three light constituents within the
approximation of exact isospin symmetry.It is worth to stress that this phenomenon
does not happen for $1^{+}$.

A simple consequence of the fact that the eigenvectors do not depend
on the masses and couplings is that the scalar nonet presents an universal
pattern of decays, the lowest state has about $55\%$ probability to decay into PP
(negligible in VV) and for the next states, in order of increasing mass: $41\%$
in VV, $41\%$ in PP and $55\%$ in VV. Identifying the lowest state of the light
nonet with the $\sigma/f_{0}(600)$ and the third one with the $f_{0}(1370)$
we get the mass of light quarks $m_{q}$ and $C_{qq}$, we find
$m_{q}\widetilde{=}351.65~MeV$ and $C_{qq}\widetilde{=}74.4~MeV$.
Notice that the quark mass and the coupling can be expressed in terms of the
masses of $\sigma$ and $f_{0}$ by:
\begin{eqnarray}
4m_{q}&=&m_{\sigma}+\left(1+\frac{17}{\sqrt{241}}\right)\frac{m_{f_{0}}-m_{\sigma}}{2} \nonumber \\
C_{qq}&=&\frac{3}{\sqrt{241}}\frac{m_{f_{0}}-m_{\sigma}}{2}. \nonumber \\
\end{eqnarray}
So it is immediate to realize that, if we would take for $m_{\sigma}$ a lower
value, around $450MeV$, as suggested by some authors,
the change in $m_{q}$ would be negligible but $C_{qq}$ would rise to $89\,MeV$

A similar determination of the parameters concerning the $s$ and $c$ quarks is
not feasible because presently we dispose only of one strange scalar as 
a possible candidate for a tetraquark ($\kappa(800)$) and none for charm.
For the $s$ quark we choose the parameters in order to reproduce the masses of the
$\kappa(800)$ as a $(qq\overline{q}\overline{s})$ state, the $a_{0}(980)$ as a
$(qs\overline{q}\overline{s})$ and the $f_{1}(1420)$ as a $1^{+}$
$(qs\overline{q}\overline{s})$ state, getting $m_{s}\widetilde{=}455.21~MeV$,
$C_{qs}\widetilde{=}58.04~MeV$ and $C_{ss}\widetilde{=}43.2MeV$.

It is quite unexpected the almost exact agreement with the parameters of our
previous calculation for the pentaquarks~\cite{ABFRT}, where we found:
$m_{q}\widetilde{=}346.8\,MeV$, $C_{qq}\widetilde{=}74.\,MeV$,
$m_{s}\widetilde{=}480\,MeV$ and for $C_{qs}$ and $C_{ss}$ we assumed the hyperfine
prescription $\frac{C_{qs}}{C_{qq}}=\frac{C_{ss}}{C_{qs}}=\frac{m_{q}}{m_{s}}$
which, as a matter of fact, is also well satisfied by the tetraquark determinations.

The parameters related to charm have been obtained requiring agreement with the
masses of the following states: $X(3872)$ as a $1^{+}$ $(qc\overline{q}\overline{c})$
state, the pair $D_{s}(2317)$ and $D_{s}(2573)$ as $0^{+}$ $(qc\overline{q}\overline{s})$
states and finally $D_{s}(2460)$ as a $1^{+}$ $(qc\overline{q}\overline{s})$ state.
The values obtained for the parameters are: $m_{c}\widetilde{=}1631\,MeV$, $C_{qc}=26\,MeV$,
$C_{cc}=18\,MeV$, $C_{sc}=17.6\,MeV$.
A direct determination from the $J/\psi$ and $\eta_{c}$ masses gives
$m_{c}\simeq 1534\,MeV$ , $C_{cc}=21.6\,MeV$. Since we expect a bigger
kinetic energy for the tetraquark together with a broader wave function, the
discrepancy goes in the right direction~\protect\footnote{The hyperfine law $\simeq 1/m_{i}m_{j}$ 
does not apply to the charm sector, since the wave function, due to
a much higher mass, is much peaked around the origin, partially compensating the
mass powers in the denominator. Actually, recent data on the $\eta_{b}$ suggest
a mass splitting with the $\Upsilon$ of the same order of the $\eta_{c}-\psi$,
and not a factor $(m_{c}/m_{b})^{2}$ smaller~\cite{ETAb}.}. On the other side if we 
determine $C_{qc}$ from the $D^{\ast}-D\ $ mass splitting, we
get $C_{qc}=26.2MeV$, in excellent agreement with the determination via the
tetraquaks spectrum.

Here it is interesting to notice that the system $Q\overline{q}$
should obey some general property as a consequence that the recoil of $Q$
can be safely neglected. So it should not depend on the mass of $Q$, but
only on the radial and orbital quantum numbers of $\overline{q}$. Since
$\overline{q}$ is very light the system would have a spatial extension that
falls in the region of dominance of the linear part of the confinement
potential, (phenomenological analysis demonstrate that the $c\overline{c}$
system falls in the logarithmic dominated region) for which well known
scaling laws~\cite{QuiggRossner} prescribe that the wave function at the
origin does not depend on the $Q$ mass, so we should expect the product
$m_{Q}\,C_{qQ}$ to be constant. A law equivalent to the constancy of
the product $m_{Q}\,C_{qQ}$ has been inferred some time ago in ref.~\cite{Gatto}
and verified for a great number of states involving charm or beauty.
\begin{widetext}
\begin{center}
\begin{table}[h]
\begin{tabular}[c]{||c||c||c||c||c||c||c||}\hline\hline
$J^{P}$ & $qq\overline{q}\overline{q}$ & $qq\overline{q}\overline{s}$ &
$qq\overline{q}\overline{c}$ & $qs\overline{s}\overline{s}$ & $ss\overline
{s}\overline{s}$ & $Decays$\\\hline\hline
$0^{+}$ & $600.^{(\ast)}\ I=0$ & $792.3^{(\ast)}\ I=1/2$ & $2141.7$ & $-$ &
$-$ & $0.55(PP);\ 1.7 10^{-3}(VV)$\\\hline\hline
$Exp$ & $f_{0}(600)$ & $\kappa(800)$ &  &  &  & \\\hline\hline
$0^{+}$ & $1046.4\ I=0,1,2$ & $1189.6\ I=1/2,3/2$ & $2442.9$ & $1472.2$ & $1611.6$%
& $0.41(PP);\ 3.1 10^{-2}(VV)$\\\hline\hline
$0^{+}$ & $1370.^{(\ast)\ }I=0$ & $1477.6\ I=1/2$ & $2661.2$ & $-$ & $-$ &
$3.1 10^{-2}(PP);\ 0.41(VV)$\\\hline\hline
$Exp$ & $f_{0}(1370)$ &  &  &  &  & \\\hline\hline
$0^{+}$ & $1816.4\ I=0,1,2$ & $1874.9\ I=1/2,3/2$ & $2962.4$ & $1996.1$ & $2058.9$%
& $1.7 10^{-3}(PP);\ 0.55(VV)$\\\hline\hline
$2^{+}$ & $1605.$\ twice $ I=0 $ and $I=1,2$ & $1686.7 I=1/2,3/2$ & $2819.8$ & $1852.3$ & $1936.1$ &
$0.5(VV);\ 0.5$ (light\ mesons)\\\hline\hline
$Exp$ & $X(1600)\ I=2$\cite{Albrecht1991F} &  &  &  & $f_{2}(2010)?$%
\cite{Etkin:1987rj} & \\\hline\hline
\end{tabular}
\caption{\label{Tab1} $0^+$ and $2^+$ states with 3 light (strange) quarks calculated
exactly according to section IV. Values of masses used in the fit are
distinguished with a (*). Experimental results, when available, are displayed
in the next row, numbers in square brackets give the reference to the experimental data.
Pauli principle fixes the isospins of the various states,so $qq\overline{q}\overline{c}$
have the same isospins as $qq\overline{q}\overline{s}$ while $qs\overline{s}\overline{s}$
have $I=1/2$ and $ss\overline{s}\overline{s}$ $I=0$. The states forbidden by the Pauli
principle are indicated by~($-$). Masses are given in MeV.}
\end{table}
\end{center}
\end{widetext}
In the case of a ``neutral'' state $(qq^{\prime}\overline{q}\overline{q}^{\prime})$,
as for hidden strangeness or charm, the $1^{+}$ CM matrix in the $\beta$-basis
is block diagonal, with a $2\times2$ block corresponding to $C=+$ and the other
$4\times4$ block to $C=-$.
So, independently of the parameters, we have two exact eigenvectors, one along the
direction $\beta_{3}\ $ (pair of color singlet vectors) and the other along $\beta_{6}$
(pair of color octet vectors). On the other hand all scalars and tensors have the
same charge conjugation, $C=+$.

It is immediate to calculate the masses of the two $C$-even states: the first
has mass $2m_{q}+2m_{q^{\prime}}+4/3(C_{qq}+C_{q^{\prime}q^{\prime}})$ and the
second $2m_{q}+2m_{q^{\prime}}-1/6(C_{qq}+18\ C_{qq^{\prime}}+C_{q^{\prime}q^{\prime}})$.
We can also calculate exactly the $2^{+}$ sector getting for
the mass $2m_{q}+2m_{q^{\prime}}+4/3(C_{qq}+C_{q^{\prime}q^{\prime}})$, the
corresponding eigenvector being along $\gamma_{1}$ (pair of color singlet vectors),
the value of the other mass is $2m_{q}+2m_{q^{\prime}}-1/6(C_{qq}-18\ C_{qq^{\prime}}+C_{q^{\prime}q^{\prime}})$
corresponding to $\gamma_{2}$ (pair of color octet vectors). A general trend
for this case is that the highest $1^{++}$ state is degenerate with the highest
$2^{+}$ state, both decaying exclusively into $V_{q\overline{q}}V_{q^{\prime}\overline{q}^{\prime}}$.
The other $1^{++}$ is below the light tensor state and has dominant decay into
$P_{q^{\prime}\overline{q}}V_{\overline{q}^{\prime}q}+P_{\overline{q}^{\prime}q}V_{q^{\prime}\overline{q}}$
while the light tensor decays into $V_{q^{\prime}\overline{q}}V_{\overline{q}^{\prime}q}$.
The states $0^{++}$ and $1^{+-}$ have to be calculated numerically, with the exception
of the case $q=q^{\prime}$, when the spectrum of the $1^{+}$ becomes highly degenerate.
In such a case, the $C$-even state $\beta_{6}$ is paired with a $C$-odd state with
eigenvector $\chi_{6}=2/3(-1,1,0,1/(2\sqrt{2}),-1/(2\sqrt{2}),0)$, the other $C$-even state
$\beta_{3}$ becomes degenerate with the $C$-odd state with eigenvector
$\chi_{3}=2/3(-1/(2\sqrt{2}),1/(2\sqrt{2}),0,-1,1,0)$. As can be seen from the table
below, the mass region $1100-1950\,MeV$ could seem to be populated by some
controversial peaks with no definite spin or $C$-parity, due to states overlapping.
\begin{widetext}
\begin{center}
\begin{table}[h]
\begin{tabular}
[c]{||c||c||c||c||c||c||c||}\hline\hline
$C$ & $-$ & $+$ & $-$ & $-$ & $+$ & $-$\\\hline\hline
$qq\overline{q}\overline{q}$ & $1109.\ I=0$ & $1158.6\ I=1$ & $1158.6\ I=1$ &
$1406.6\ I=0,1,2$ & $1605.\ I=1$ & $1605.\ I=1$\\\hline\hline
$ss\overline{s}\overline{s}$ & $-$ & $-$ & $-$ & $1820.8$ & $-$ &
$-$\\\hline\hline
$Decays$ & $PV$ &$PV$& $PV$ & $PV$ & $VV$ & $VV$\\\hline\hline
\end{tabular}
\caption{\label{Tab1} Axial states made of all light (in the limit of exact isospin)
or strange (anti) quarks calculated exactly,
according to section IV. They have definite charge conjugation. The states forbidden by
the Pauli principle are indicated by ($-$). Masses are given in MeV.}
\end{table}
\end{center}
%
Even if no candidates have been observed let us, for completeness, give the spectrum
of strange and charmed axials:
\begin{center}
\begin {table}[h]
\begin{tabular}[c]{||l||l||l||l||l||l||l||}\hline\hline
$qc\overline{q}\overline{q}$ & $2329.3$ & $2515.6$ & $2611.7$ & $2727.8$ &
$2785.8$ & $2877.73$\\\hline\hline
$Decays$ & $0.55(\pi,\eta)D^{\ast}$ & $0.39(\pi,\eta)D^{\ast}$ & $0.47(\omega,\rho)D$
& $0.28(\omega,\rho)D$ & $0.47(\omega,\rho)D^{\ast}$
 & $0.46(\omega,\rho)D^{\ast}$ \\
&&&&& $0.11(\omega,\rho)D$ &
\\\hline\hline
$qq\overline{q}\overline{s}$ & $1207.5$ & $1302.7$ & $1308.6$ & $1513.4$ &
$1672.7$ & $1703.$\\\hline\hline
$Decays$ & $0.56(\pi,\eta)K^{\ast}$  & $0.17(\omega,\rho)K$
&$0.52(\omega,\rho)K$ & $0.21(\omega,\rho)K$&$0.50(\omega,\rho)K^{\ast}$ & $0.50(\omega,\rho)K^{\ast}$ \\
&&$0.28(\pi,\eta)K^{\ast}$&&$0.12(\pi,\eta)K^{\ast}$&&
\\\hline\hline
Isospin & $1/2$ & $1/2,3/2$ & $1/2$ & $1/2,3/2$ & $1/2$ & $1/2,3/2$%
\\\hline\hline
\end{tabular}
\caption{\label{Tab1} Charmed and strange axial mesons, calculated numerically.
Masses are in MeV. The non negligible decay channels are indicated in the row
below.}
\end{table}
\end{center}
\end{widetext}
When an object contains a pair of (anti) quarks, Pauli principle implies the
absence of some states or, otherwise, if the pair is made of light quarks,
restrictions on the isospin content, according to the correspondence
$I=0\rightarrow 21_{cs}$ and $I=1\rightarrow15_{cs}$. This has been taken
into account in the elaboration of Tables I, II, III, where
Pauli forbidden states are indicated by a hyphen.
The very interesting cases of hidden strangeness/charm and tetraquarks
with $C=\pm~S=1$ were calculated numerically and are given in the
Table~\ref{tavolona}. The interest for the somewhat chimerical states with
$C=-~S=1$ and $C=2$ i.e. of kind $(cs\overline{q}\overline{q})$ and
$(cc\overline{q}\overline{q})$, is justified by the fact that they provide
a clear signature for tetraquarks. In the case of $I=0$ the first decays
into $D^{+}{K}^{-}$ and $D^{0}\overline{K}^{0}$ and the second into $D^{+}D^{0}$.
Since in both cases the objects carrying strangeness or charm are necessarily
a pair of quarks and obviously they cannot form by themselves color singlets,
so the occurrence of such states is possible only if the pair of quarks combine
with at least a pair of antiquarks.

\section{Discussion on the results for tetraquarks}

First of all, let us recall that the information we used in the fit
involves only the mass spectrum, so the pattern of decays may be considered
as ``predictions''. Let us cite the observed dominance of $\pi\pi$ in the
$f_{0}(600)$ decay and of $\rho\rho$ in that of
$f_{0}(1370)$\cite{PDG}\cite{Gasp}, the dominance of the $\pi K$ channel for
$\kappa(800)$ (unfortunately, by now, omitted from PDG).

For the axials we obtained the dominance of $\overline{K}K^{\ast}+cc$
($KK\pi$ probably arising from a off-shell $K^{\ast})$ for the $f_{1}(1420)$
and, analogously, the dominance of $\overline{D}D^{\ast}+cc$ for the $X(3872)$.
\newpage
\begin{widetext}
\begin{center}
\begin{table}[h]\label{tavolona}
\begin{tabular}{||c||c||c||c||c||c||c||}\hline 
$J^{P}$ & $qs\overline{q}\overline{s}$ & $cs\overline{q}\overline{q}$ &
$qc\overline{q}\overline{s}$ & $qc\overline{q}\overline{c}$ & $cc\overline
{q}\overline{q}$ & $cs\overline{c}\overline{s}$\\\hline\hline
$0^{+}$ & $981.^{(\ast)}$ & $2326.7$ & $2315.^{(\ast)}$ & $3562.7$ & $3643.1$%
& $3904.5$\\\hline\hline
$Exp$ & $a_{0}(980)$ &  & $D_{s_{0}}^{\ast\pm}(2317)$ &  &  & \\\hline\hline
$0^{+}$ & $1330.3$ & $2592.3$ & $2574.^{^{(\ast)}}$ & $3799.3$ & $3870.8$ &
$4060.8$\\\hline\hline
$Exp$ & $a_{0}(Y)$\cite{a01330} &  & $D_{s_{1}}^{\pm}(2573)$ &  &  &
\\\hline\hline
$0^{+}$ & $1586.1$ & $2757.6$ & $2773.7$ & $3979.3$ & $3898.6$ &
$4181.$\\\hline\hline
$0^{+}$ & $1934.9$ & $3028.$ & $3028.4$ & $4148.4$ & $4144.5$ & $4295.$%
\\\hline\hline
$Exp$ &  &  &  &  &  & $X(4350)$\cite{X(4350)}\\\hline\hline
$1^{+}$ & $1327.6$ & $2503.7$ & $2469.3^{^{(\ast)}}$ & $3682.9$ & $3795.3$ &
$4016.41$\\\hline\hline
$Exp$ &  &  & $D_{s_{1}}^{\pm}(2460)$ &  &  & \\\hline\hline
$1^{+}$ & $1420.^{(\ast)}$ & $2674.5$ & $2634.6$ & $3871.9^{(\ast)}$ &
$3847.8$ & $4109.4$\\\hline\hline
$Exp$ & $f_{1}(1420)$ &  &  & $X(3872)$ &  & $Y(4140)$\cite{Y(4140)}
\\\hline\hline
$1^{+}$ & $1461.$ & $2692.$ & $2736.4$ & $3924.6$ & $3927.8$ & $4132.1$%
\\\hline\hline
$1^{+}$ & $1618.9$ & $2822.8$ & $2823.$ & $3980.5$ & $3991.6$ & $4172.5$%
\\\hline\hline
$1^{+}$ & $1770.5$ & $2857.8$ & $2889.3$ & $4057.5$ & $3992.2$ &
$4225.9$\\\hline\hline
$1^{+}$ & $1773.$ & $2959.5$ & $2951.$ & $4088.5$ & $4084.78$ & $4254.$%
\\\hline\hline
$2^{+}$ & $1768.2$ & $2889.$ & $2900.2$ & $4027.9$ & $4021.2$ & $4215.$%
\\\hline\hline
$2^{+}$ & $1770.5$ & $2906.9$ & $2912.2$ & $4088.5$ & $4061.6$ &
$4254.$\\\hline
\end{tabular}
\caption{\label{Tab1} Spectrum of the tetraquarks calculated numerically.
States used in the fit are marked with a (*). When experimental data are 
available they are displayed in the next row, reference to the source are
given in square brackets. Masses are in Mev.}
\end{table}
\end{center}
\end{widetext}
Since they are pure $\beta_{6}$ states these channels are exclusive.
In particular, for $X(3872)$, the observed decays into 
$\rho(\omega)J/\psi$, can be explained by one gluon exchange in the t-channel,
since those rates are comparable with the process being $O(\alpha_{s})$.
For $D_{s_{0}}^{\ast\pm}(2317)$ the only kinematically allowed open door channel
is $\pi^{0} D_{s}^{\pm}$, it is just below the $DK$ threshold, at $2359\,MeV$.
The relevant components are $\alpha_{1}=0.78$ , $\epsilon_{1}=0.70$, so predicting
strong dominance of the $\pi^{0} D_{s}^{\pm}$ decay. In the case of
$D_{s_{2}}^{\ast\pm}(2573)$ that we interpreted to be $0^{+}$ (even if it is
also consistent with a $2^{+}$) the only observed decay is $D^{0}K^{\pm}$ while
$D^{0\ast}(2007)K^{\pm}$ is not, so in agreement with PP prescription arising from
scalar nature of the state. Nevertheless, also in this case the components
are almost equal $\alpha_{1}=0.60$ ($\pi^{0} D_{s}^{\pm}$),
$\epsilon_{1}=0.68$ ( $D^{0}K^{\pm}$) and so we could expect the
$\pi ^{0}D_{s}^{\pm}$ to be relevant, as well. Experimental data neither confirm
nor disprove this point. Finally the axial state $D_{s_{1}}^{\pm}(2460)$,
that we put at $2469.3\,MeV$, has a large component along $\beta_{1}(0.87)$,
which corresponds to the dominant $\pi^{0}D_{s}^{\pm\ast}$ channel.
The $\omega D^{\pm}_{s}$ decay (notice the state $D_{s_{1}}^{\pm}(2460)$
has $I=0$) has a tiny component $\beta_{2}=0.024$ and is also kinematically
inaccessible. It remains to explain the large branching fraction in
$D^{\pm}_{s}\gamma$, suggesting that the state is very narrow, albeit the
experimental upper bound is not much restrictive, $\Gamma\leq3.5MeV$.

Concerning the two degenerate states, the isoscalar $f_{0}$ and the isovector
$a_{0}$, at $980MeV$, they can only decay into $\eta\pi$ and $K\overline{K}$,
other channels being too high. We predict $\alpha_{1}=0.75,~\epsilon_{1}=0.74$,
and if we take the corrections for the mixing $\eta_{0}-\eta_{8}$ with a mixing
angle $\theta=-16^{\circ}$ (as obtained recently in $\gamma\gamma\rightarrow X$),
we find for the ratio of
$g_{a_{0}K\overline{K}}^{2}/g_{a_{0}\eta\pi}^{2}\widetilde{=}2.48$,
to be compared with the value recently obtained by the KLOE experiment~\cite{KLOE}
of $0.67\pm0.06\pm0.13$. This abnormally large coupling for $\eta\pi$ cannot be
obtained by chromo-magnetism alone, it has been explained recently~\cite{Hooft}
by non perturbative effects induced by instantons. Analogously for the dominant
decay $f_{0}\rightarrow\pi\pi$, which violates OZI rule, we have to rely on the
above solution, in association with $f_{0}(980)-\sigma$ mixing.

We predict a companion (which is a mixture of $8_ {F}$ and  $27_ {F}$), for the $a_{0}(980)$
at $1330.3MeV$ coupled to $\eta\pi$, $\eta^{\prime}\pi$ and $K\overline{K}$.
It was recently observed~\cite{a01330} in $\gamma\gamma\rightarrow\eta\pi^{0}$
and named $a_{0}(Y)$ with an observed mass of $1316\pm25MeV$.

In the hidden charm-strange sector ($cs\overline{c}\overline{s}$)we have found
two candidates for newly discovered states. The first is the pure $\beta_{6}$
$1^{++}$ state at $4109.4MeV$ which we propose to identify to the narrow
state $Y(4140)$ found at CDF \cite{Y(4140)} in $B^{+}\rightarrow
XK^{+},X\rightarrow J/\psi\phi,$ with a mass$\ 4143\pm2.9\pm1.2MeV$ and a
width of $11.7_{-5.0}^{+8.3}\pm3.7MeV$. As the $X(3872)$ the later has
dominant decays into $\overline{D_{s}}D_{s}^{\ast}+cc$ (threshold at
$4080MeV$), but can also decay into $J/\psi\phi$ ( threshold at $4116.4MeV$).
The choice of spin one\footnote{The interpretation of the $Y(4140)$ as an
axial was already contemplated in ref \cite{Stancu}, albeit not excluding the
$0^{++}$ alternative.} is strongly suggested by
the fact that it was not observed in $\gamma\gamma\rightarrow X$ by
BELLE\cite{X(4350)}. The second state is a $0^{++}$ at $4295MeV$ , with
predominant decays into $J/\psi\phi$ ($\alpha_{2}\simeq0.81$) and
$D_{s}^{\ast}\overline{D_{s}^{\ast}}$ ($\beta_{2}\simeq0.69$), to be interpreted as
the $X(4350),$ discovered by BELLE in the same experiment\cite{X(4350)}, with a mass
$4350.6_{-5.1}^{+4.6}\pm0.7MeV$ and width $13.3_{-9.1}^{+17.9}\pm4.1MeV$.
Taking into account phase space, we find the $J/\psi\phi$ channel to be twice
more probable than the $D_{s}^{\ast}\overline{D_{s}^{\ast}}$ one.

Among the non well established states there is a $2^{+}$ state $X(1600)$
(with I=2)~\cite{Albrecht1991F} at $1600\pm100MeV$ that, if interpreted as
$(qq\overline{q}\overline{q})$, is compatible with our predictions and,
according to the previous section, it has to be degenerate with the highest
$1^{++}$, the later being possibly hidden by some ($L=1\ q \overline{q}$)
state of the $a_{1}$ family.

It is not excluded that presently we have already seen some
$(ss\overline{s}\overline{s})$ states, one of these could be the $f_{0}(2010)$
found around $2011\pm70MeV$~\cite{Etkin:1987rj} that is identifiable to our
$2^{+}$ state at $1936\,MeV$.
We predict a $1^{+},$ $(qs\overline{q}\overline{s})$ state, with a mass of $1327.6MeV$
decaying predominantly into $\pi\phi$ ($\beta_{1}\backsimeq0.91$) and another one at
$1773MeV$ with important components along $\beta_{1}\backsimeq0.34$
($\eta_{s}V$) and $\beta_{2}\backsimeq0.15$ ($\pi\phi$, $\eta\phi$),while
$\chi_{3}$ is also very large, the state is below threshold for
$K^{\ast}\overline{K}^{\ast}$. The later could be, possibly identified to the $X(1835)$
found at BES ~\cite{BESgammaX} at $1834\pm6\,MeV$ and width
$67.7\pm20.3\pm7.7\,MeV$, decaying into $\pi^{+}\pi^{-}\eta^{\prime}$.
The spin-parity of the $X(1835)$ is not known and it was, initially, supposed
to be related to a $p\bar{p}$ threshold enhancement, due to the strong
dominance of the channel $\pi^{+}\pi^{-}\eta^{\prime}$.
 
We also predict a $0^{+}$ $ss\bar{s}\bar{s}$ state at $2058.9 MeV$, is strongly
coupled to $\phi\phi$, so it would arise as a $\phi\phi$ threshold enhancement.

\section{Negative parity states built with three quarks and three antiquarks}

Today it seems to exist experimental evidence for the occurrence of
baryon-antibaryon states. People could have the tendency to interpret them as
molecular states, but as said before, there is no clear distinction between
chromo-magnetism and the molecular point of view as long as we do not neglect
some configurations of the diquarks. In obtaining the predictions of
chromo-magnetism, since the number of candidates is not enough to completely
determine the parameters, we will tentatively assume for the masses and
chromo-magnetic couplings of the quarks in the baryon-antibaryon system 
the same as for tetraquarks. As mentioned before, masses could be larger
due to the fact that they are defined including the kinetic energy.
On the other hand, couplings could be smaller mainly because the wave
function is more spread.

A complete calculation is very complex and probably not of immediate utility
in view of the scarcity of these states. We treat two cases, the first is
related to $p\bar{p}$ states and concerns $(qqq\overline{q}\overline{q}\overline{q})$
systems, the second deals with the production of a variety of states of the kind
$(qqq\overline{q}\overline{q}\overline{Q})$ or
$(qqQ\overline{q}\overline{q}\overline{Q})$, where $Q$ denotes an $s$ or $c$ quark.

It is natural to work with what we call the baryon-antibaryon basis. In the
first case, since we are interested in a $p\bar{p}$ pair, it is enough to take the
sub block $qqq$ in the $70$ of $SU(6)_{cs}$ (and $\overline{q}\overline{q}\overline{q}$
in the $\overline{70}$). The decomposition of the $70$, under $SU(3)_{c}\otimes SU(2)_{s}$
is given by: $70_{cs}=(8_{c},4_{s})+(8_{c},2_{s})+(10_{c},2_{s})+(1_{c},2_{s})$.
We can construct 4 color singlets of spin 0 and 6 of spin 1, which are below:
\begin{center}
\textbf{Spin 0}
\end{center}
\begin{center}
\begin{tabular}{ll}
$\left|1\right\rangle = [(1_c,2_s),(1_c,2_s)];  $&$ \left|2\right\rangle = [(8_c,2_s),(8_c,2_s)]; $ \\
$ $ \\
$\left|3\right\rangle = [(8_c,4_s),(8_c,4_s)];  $&$ \left|4\right\rangle = [(10_c,2_s),(\overline{10}_c,2_s)] $ \\
\end{tabular}
\begin{equation} \label{eqs0:6}
\end{equation}
\end{center}
\begin{center}
\textbf{Spin 1}
\end{center}
\begin{center}
\begin{tabular}{ll}
$\left|1\right\rangle =[(1_c,2_s),(1_c,2_s)]; $&$ \left|2\right\rangle =[(8_c,2_s),(8_c,2_s)];  $ \\
$ $ \\
$\left|3\right\rangle =[(8_c,4_s),(8_c,4_s)]; $&$ \left|4\right\rangle =[(10_c,2_s),(\overline{10}_c,2_s)]; $ \\
$ $ \\
$\left|5\right\rangle =[(8_c,2_s),(8_c,4_s)]; $&$ \left|6\right\rangle =[(8_c,4_s),(8_c,2_s)] $ \\
\end{tabular}
\begin{equation} \label{eqs1:6}
\end{equation}
\end{center}
Evaluating the chromo-magnetic operator of Eq.~[\ref{eq:HCM}] between these states
we get the 2 matrices, describing chromo-magnetism in the 2 sectors, given in
Eqs.~[\ref{nps0},\ref{nps1}], where it was assumed the same ordering as above.

This has been done using a computer, but since we are in fact in the symmetry limit,
it can also be calculated by purely group theoretical means. It furnishes a valuable
check of the machine's symbolic calculation. It is straightforward to obtain the expression in
terms of Casimir operators:
{\small
\begin{eqnarray}
O_{CM\text{ }}&=&[ C_{6}(R_{3q})+C_{6}(R_{3\overline{q}}
)-\frac{1}{2}C_{3}(R_{3q})-\frac{1}{2}C_{3}(R_{3\overline{q}})\nonumber\\
&&-\frac{1}{3}S_{3\overline{q}}(S_{3\overline{q}}+1)-\frac{1}{3}S_{3\overline
{q}}(S_{3\overline{q}}+1)-12]\nonumber\\
&&-[C_{6}(H)-C_{6}(R_{3q})-C_{6}(R_{3\overline{q}})+\frac{1}{2}C_{3}(R_{3q})
\nonumber\\
&&+\frac{1}{2}C_{3}(R_{3\overline{q}}) -\frac{1}{3}S_{H}(S_{H}+1) \nonumber\\
&& +\frac{1}{3}S_{3q}(S_{3q}+1)
+\frac{1}{3}S_{3\overline{q}}(S_{3\overline{q}}+1)]
\end{eqnarray}
}
where $H$ stands for the representation of the hexaquark in $SU(6)_{cs}$, with
$S_{H}$ being its spin (0 or 1 in the present case), $R_{3q}$ and
$R_{3\overline{q}}$ the representations of the 3 quarks and 3 antiquarks
subsystems, respectively (of both groups , $SU(6)_{cs}$ and $SU(3)_{c})$,
$S_{3q}$ and $S_{3\overline{q}}$ being their spins. As before, $C_{6}$ and
$C_{3}$ are the quadratic Casimir operators of $SU(6)_{cs}$ and $SU(3)_{c}$.
In the first square brackets we have isolated the contribution of the quark-quark and
antiquark-antiquark interactions, while in the second the contribution for
quark-antiquark interactions. Here a severe complication arises: the Casimir
operators in the second bracket are not diagonal. As the operator $O_{CM}$
transforms as the $35$ of $SU(6)_{cs}$, it does not leave the $70$ and, thus the
Casimir operators present in the first bracket are diagonal, while for the
second one, representation mixing remains possible and in fact it occurs.

The hexaquark state $(qqq\overline{q}\overline{q}\overline{q})$, we have
designated by $H$, transforms under $SU(6)_{cs}$ as one of irreducible
representations (or mixings thereof) arising in the product below:
$70\otimes\overline{70}=1+35_{1}+35_{2}+189+280+\overline{280}+405+3675$.
For $0^{-}$ we have to select the blocks that contain components 
transforming as $(1_{c},1_{s})$, and for the $1^{-}$ as $(1_{c},3_{s})$. It is
indicated below the relevant representations and the number of components of
the suitable color singlets contained in each one:
\begin{center}
\begin{tabular}{lll}
$0^{-}: $&$ (1_{c},1_{s})\subset $&$ 1;~189(1);~405(1);~3675(1)$ \\
$ $ \\
$1^{-}: $&$ (1_{c},3_{s})\subset $&$ 35_{1}(1);~35_{2}(1);~280(1);~\overline{280}(1);3675(2)$. \\
\end{tabular}
\end{center}
The matrix elements were found through the determination of the appropriate
Clebsch Gordan coefficients for the above decomposition.

Let us now consider states of the kind $(qqQ\overline{q}\overline{q}\overline{Q})$
($Q$ being an $s$ or $c$ quark), for which some experimental evidence is available.
The Pauli principle implies that the pair of light (anti-)quarks in the (anti-)baryonic
block $qqQ$ ($\overline{q}\overline{q}\overline{Q}$) must transform under $SU(6)_{cs}$
as a $21_{cs}$ ($\overline{21}_{cs}$) for $I=0$ and as a $15_{cs}$ ($\overline{15}_{cs}$)
in the case of $I=1$. States such $(qq)_{21_{cs}}Q(\overline{q}\overline{q})_{(\overline{21}_{cs})}\overline{Q}$
have I=0 and are relevant for the $\Lambda\overline{\Lambda}$
($\Lambda_{c}\overline{\Lambda}_{c}$) channels. For shortness, we shall call them
the $(21,\overline{21})$ basis. The other case, namely $(qq)_{15_{cs}}Q(\overline{q}
\overline{q})_{(\overline{15}_{cs})}\overline{Q}$ is the $(15,\overline{15})$ basis
and comprises hexaquarks with $I=0,1,2$. This base will be used in the calculation of the
$\Sigma\overline{\Sigma}$ channel.

A criterion to build the physical states, i.e. the color singlets of the six
quark system, is to combine successively $qq$ with $Q$ (and analogously for
the antiquarks) in all possible ways regarding the color group $SU(3)_{c}$
and then combining with those of the antiquarks. This can be easily done
using the decompositions of $SU(6)_{cs}\rightarrow SU(3)_{c}\otimes SU(2)_{s}$:
$21_{cs}=(\overline{3}_{c},1_{s})+(6_{c},3_{s})$ and $15_{cs}=(6_{c},1_{s})+(\overline{3}_{c},3_{s})$.
Taking into account the genealogy of the states, we get for each basis, a total of $14$
color singlets. They are displayed below~\footnote{The representation
$8_{sim}$ is the color octet symmetric under the exchange of the colors
of the light quark pair.}.
The convention we use is the following: the composition
of the baryonic ($qqQ$) with anti-baryonic blocks ($\overline{q}\overline{q}\overline{Q}$)
is indicated by a (*), each block is
enclosed by a square bracket and within each bracket we placed on the left the
color-spin content of ($qq$) followed by that
of $Q$ \ (and analogously for the antiquarks).

As will be seen in the next section, we have also interest to build the basis
for the system $\Lambda_{c}\overline{p}$.We use the ordering convention
$(\overline{q}_{1}\overline{q}_{2}\overline{q}_{3}q_{4}q_{5}c_{6})$.
The $\overline{p}$, as previously, is put in a $\overline{70}_{\beta}$
(antisymmetric in 1,2) and the $\Lambda_{c}$ (as the Pauli antisymmetry applies only to the
pair 4 and 5) in a ${70}_{\alpha}$ (symmetric with respect to 4 and 6) and a $56$,
which decomposes under $SU(3)_{c}\otimes SU(2)_{s}$ as: $(10,4)+(8,2)$. The mandatory
anti-symmetrization with respect to flavor of the pair 4 and 5 implies isospin $0$ for the
$\Lambda_{c}$.
\begin{center}
Basis $(21,\overline{21})$ for spin 1
\end{center}
\begin{center}
\begin{tabular}{ll}
$\mathbf{[(\overline{3},1)(3,2)] \ast [(3,1)(\overline{3},2)]}
\Rrightarrow $&$ \left|1\right\rangle =(1,2) \ast (1,2) $ \\
              &$ \left|2\right\rangle =(8,2) \ast (8,2) $ \\
$ $ \\
$\mathbf{[(6,3)(3,2)] \ast [(3,1)(\overline{3},2)]}
\Rrightarrow $&$ \left|3\right\rangle =(8_{sim},4) \ast (8,2) $ \\
              &$ \left|4\right\rangle =(8_{sim},2) \ast (8,2) $ \\
$ $ \\
$\mathbf{[(\overline{3},1)(3,2)] \ast [(\overline{6},3)(\overline{3},2)]}
\Rrightarrow $&$ \left|5\right\rangle=(8,2) \ast (8_{sim},4)  $ \\
              &$ \left|6\right\rangle=(8,2) \ast (8_{sim},2)  $ \\
$ $ \\
$\mathbf{[(6,3)(3,2)] \ast [(\overline{6},3)(\overline{3},2)]}
\Rrightarrow $&$ \left|7\right\rangle=(8_{sim},4) \ast (8_{sim},4) $ \\
              &$ \left|8\right\rangle=(8_{sim},4) \ast (8_{sim},2) $ \\
              &$ \left|9\right\rangle=(8_{sim},2) \ast (8_{sim},4) $ \\
              &$ \left|10\right\rangle=(8_{sim},2) \ast (8_{sim},2)  $ \\
              &$ \left|11\right\rangle=(10,4) \ast (\overline{10},4)  $ \\
              &$ \left|12\right\rangle=(10,4) \ast (\overline{10},2)  $ \\
              &$ \left|13\right\rangle=(10,2) \ast (\overline{10},4)  $ \\
              &$ \left|14\right\rangle=(10,2) \ast (\overline{10},2)  $ \\
\end{tabular}
\begin{equation}
\end{equation}
\end{center}
\begin{center}
\newpage
Basis $(15,\overline{15})$ for spin 1
\end{center}
\begin{center}
\begin{tabular}{ll}
$\mathbf{[(\overline{3},3)(3,2)] \ast [(3,3)(\overline{3},2)]}
\Rrightarrow   $&$ \left|1\right\rangle=(1,4) \ast (1,4)  $ \\
                &$ \left|2\right\rangle=(1,4) \ast (1,2)  $ \\
                &$ \left|3\right\rangle=(1,2) \ast (1,4)  $ \\
                &$ \left|4\right\rangle=(1,2) \ast (1,2)  $ \\
                &$ \left|5\right\rangle=(8,4) \ast (8,4)  $ \\
                &$ \left|6\right\rangle=(8,4) \ast (8,2)  $ \\
                &$ \left|7\right\rangle=(8,2) \ast (8,4)  $ \\
                &$ \left|8\right\rangle=(8,2) \ast (8,2)  $ \\
$ $ \\
$\mathbf{[(\overline{3},3)(3,2)] \ast [(\overline{6},1)(\overline{3},2)]}
\Rrightarrow   $&$ \left| 9\right\rangle=(8,4) \ast (8_{sim},2)   $ \\
                &$ \left|10\right\rangle=(8,2) \ast (8_{sim},2)   $ \\
$ $ \\
$\mathbf{[(6,1)(3,2)] \ast [(3,3)(\overline{3},2)]}
\Rrightarrow   $&$ \left|11\right\rangle=(8_{sim},2) \ast (8,4)   $ \\
                &$ \left|12\right\rangle=(8_{sim},2) \ast (8,2)   $ \\
$ $ \\
$\mathbf{[(6,1)(3,2)] \ast [(\overline{6},1)(\overline{3},2)]}
\Rrightarrow   $&$ \left|13\right\rangle=(8_{sim},2) \ast (8_{sim},2)   $ \\
                &$ \left|14\right\rangle=(10,2) \ast (\overline{10},2)    $ \\
\end{tabular}
\begin{equation}
\end{equation}
\end{center}
We have 5 states for spin 0 and 9 states for spin 1, they are given below:
\begin{center}
\textbf{Spin 0}
\end{center}
\begin{center}
\begin{tabular}{ll}
$\left|1\right\rangle=(1,2)_{\beta}(1,2)_{\alpha}                 $&
$\left|2\right\rangle =(8,2)_{\beta}(8,2)_{\alpha}                $\\
$\left|3\right\rangle=(8,4)_{\beta}(8,4)_{\alpha}                 $&
$\left|4\right\rangle =(\overline{10},2)_{\beta}(10,2)_{\alpha}   $\\
$\left|5\right\rangle =(8,2)_{\beta}(8,2)_{56}                    $&\\
\end{tabular}
\begin{equation}
\end{equation}
\end{center}
\begin{center}
\textbf{Spin 1}
\end{center}
\begin{center}
\begin{tabular}{ll}
$\left|1\right\rangle =(1,2)_{\beta}(1,2)_{\alpha}                      $&
$\left|2\right\rangle =(8,2)_{\beta}(8,2)_{\alpha}                      $\\
$\left|3\right\rangle =(8,4)_{\beta}(8,4)_{\alpha}                      $&
$\left|4\right\rangle =(\overline{10},2)_{\beta}(10,2)_{\alpha}         $\\
$\left|5\right\rangle =(8,2)_{\beta}(8,4)_{\alpha}                      $&
$\left|6\right\rangle =(8,4)_{\beta}(8,2)_{\alpha}                      $\\
$\left|7\right\rangle =(8,2)_{\beta}(8,2)_{56}                          $&
$\left|8\right\rangle =(8,4)_{\beta}(8,2)_{56}                          $\\
$\left|9\right\rangle =(\overline{10},2)_{\beta}(10,4)_{56}             $&\\
\end{tabular}
\begin{equation}
\end{equation}
\end{center}
With the introduction of appropriate color and spin projectors, it is
easy to build explicitly the above basis. Symbolic expressions for  the
matrix elements of the chromo-magnetic operator $O_{CM}$ were obtained with
the help of FORM~\cite{form}. The explicit expressions for the CM matrices
for the three mentioned cases are collected in Appendix C.
It was assumed for the CM matrices the same ordering as for the above states.
The mass spectrum of the most interesting baryon-antibaryon states are 
given in appendix A.

\section{Experimental evidence for hexaquarks}

1) We predict a $0^{-}$state ($qqq\overline{q}\overline{q}\overline{q}$),
strongly coupled to the $p\bar{p}$ channel (the component along $p\bar{p}$ is $0.894$),
just below the threshold ($1876.54\,MeV$), it has a mass of $1874\,MeV$.
This is in agreement with the first observation of a narrow enhancement near $p\overline{p}$
threshold by the BES collaboration~\cite{Bai06} in $J/\psi\rightarrow p\bar{p}\gamma$,
then named $X(1859)$. Until now both the $J^{P}$ assignments $0^{+}$ or $0^{-}$
remain equally possible. It was found at a mass $m_X=1859\pm^{3}_{10}\pm^{5}_{25}\,MeV$
having a width smaller than $30~MeV$. The state we found is slightly higher,
just $7\,MeV$ above the experimental upper limit. They estimated a branching ratio of
$B(B\rightarrow\gamma X)\,B(X\rightarrow p\overline{p})\backsimeq\,7.10^{-5}$.

2) Also relevant for the light hexaquarks ($qqq\overline{q}\overline{q}\overline{q}$)
may be a quite broad $1^{-}$ enhancement above $p\bar{p}$ threshold with mass
$1935\pm20\,MeV$ and width $\Gamma=215\pm30\,MeV$ proposed about $30$ years ago~\cite{Eva}.
We have a very good candidate for this state at a mass $1911.5\,MeV$ with a large
component ($0.61$ ) along the $p\bar{p}$ channel. However here some caution is needed,
because the evidence is based on a partial wave analysis and one would have to check
if the analysis is compatible with the inclusion of the additional $0^{-}$ state just mentioned above.

3) We have also a pretty good candidate for the $Y(2175)$, a $1^{--}$ state
recently seen at the BaBar detector~\cite{BABARY} at a mass $2170\pm10\pm 15\,MeV$
(with a width $\Gamma=58\pm16\pm20\,MeV$). We predict a singly hidden strangeness
state ($qqs\overline{q}\overline{q}\overline{s}$) strongly coupled to the
$\Lambda \bar{\Lambda}$ channel (with a component of $0.6$ along this direction)
with a mass $2184\,MeV$. Since this state is below the $\Lambda \bar{\Lambda}$ threshold
(around $2231\,MeV$) it has to decay mostly into mesons. In fact BaBar observed this
state in the decay $Y\rightarrow f_{0}(980)\phi$ (through $f_{0}\rightarrow\pi\pi$).
The $Y(2175)$ has been confirmed by the BES collaboration~\cite{BES4} in
$J/\psi\rightarrow\eta f_{0}(980)\phi$ at a mass $m=2186\pm10\pm16\,MeV$ and a width
$\Gamma=65\pm23\,MeV$.

4) The peak in $\Lambda_{c}$ $\bar{p}$ seen at the mass $m=3350_{-20}^{+10}\pm29\,MeV$
and width $\Gamma=70_{-30}^{+40}\pm40\,MeV$ in $B^{-}\rightarrow\Lambda_{c}\bar{p}\pi^{-}$~\cite{Gaby}
may be identified with a $0^{-}$ strange charmed hexaquark, we predict to be at $3339\,MeV$. There is also
a $1^{-}$ at lower mass, $3274\,MeV$, with a component of the same order ($0.35$). All the states
strongly coupled to $\Lambda_{c} \bar{p}$ are below the threshold ($3225\,MeV$), on the other side
those above the threshold, with the exception of the two above mentioned states, have negligible couplings.
This implies that these two states are the only ones observable in the baryonic channel.
It is useful to remark that the experiment privileges the spin 0 assignment.

5) In the singly hidden charm sector ($qqc\overline{q}\overline{q}\overline{c}$),
 the heaviest states are loosely coupled to the
$\Lambda_{c} \overline{\Lambda}_{c}$, and the reasonably coupled states are just above or below
the threshold ($4573\,MeV$). We display these states and the value of the component
along the baryonic channel:
\begin{table}[h]
\begin{tabular}
[c]{||c||c||c||c||c||c||c||c||} \hline 
$Mass(MeV)$ & $4533$ & $4556$ & $4575$ & $4614$ & $4642$ & $4658$ &
$4670$ \\\hline\hline
$comp.\ in\ \Lambda_{c}\overline{\Lambda}_{c}$ & $0.41$ & $0.21$ & $0.52$ &
$0.42$ & $0.48$ & $0.16$ & $0.24$\\ \hline 
\end{tabular}
\end{table}
\break
\bigskip
As a matter of fact, recently, a resonance decaying into $\Lambda_{c}\bar{\Lambda}_{c}$
has been seen by the Belle detector~\cite{Y4660,DETROIT} at $m=4634_{-7-8}^{+8+5}\,MeV$ and 
$\Gamma=92_{-24-21}^{+40+20}\,MeV$ , compatible~\cite{Choi,DETROIT} with
$Y(4660)\rightarrow \psi^{\prime}\pi\pi$~\cite{PSIPIPI,DETROIT}.
Anyway the fact that the component along the baryonic channel is not 
strongly dominant is welcome, since it is opportune to leave some room for the decay into
$\psi^{\prime}\pi\pi$.
Recently it was proposed to interpret the above state as an excited L=1 tetraquark~\cite{QED}.

We have also calculated the spectrum of the singly hidden strangeness states
($qqs\overline{q}\overline{q}\overline{s}$) relevant to the $\Sigma\overline{\Sigma}$
channel, using, along the same lines, the $(15,\overline{15})$ basis. We find only two states
strongly coupled to $\Sigma\overline{\Sigma}$, both are around the threshold, $2380MeV$, one
being below threshold at a mass of $2356MeV$ the other above, at $2454MeV$. Until now, there is
no experimental evidence for these states.

\section{Conclusion}

The full chromo-magnetic Hamiltonian proved to be very effective in
providing for an unified treatment of tetraquarks and hexaquarks. Besides
reproducing the pattern of decays of currently accepted tetraquarks, it also
predicts a companion for the $a_{0}(980)$ at a mass around $1330MeV$, which has
been confirmed by experiments, as the scalar named $a_{0}(Y)$ and two
$cs\overline{c}\overline{s}$ states, the $Y(4140)$ and the $X(4350)$. A number of
candidates were compared with data for the baryon-antibaryon resonances, namely
$p \overline{p}$, $\Lambda_{c}\overline{\Lambda}_{c}$, $\Lambda_{c}\overline{p}$
quite successfully.

\section*{Acknowledgments}

We are indebted to our friends, G. D'Ambrosio and N. Lo Iudice, for useful
discussions.

\appendix

\newpage
\begin{widetext}
\section{Spectrum of the $B\overline{B}$ states}
\begin{center}
\begin{figure*}[h]
\includegraphics[bb=0 0 842 596,scale=.7,angle=90]{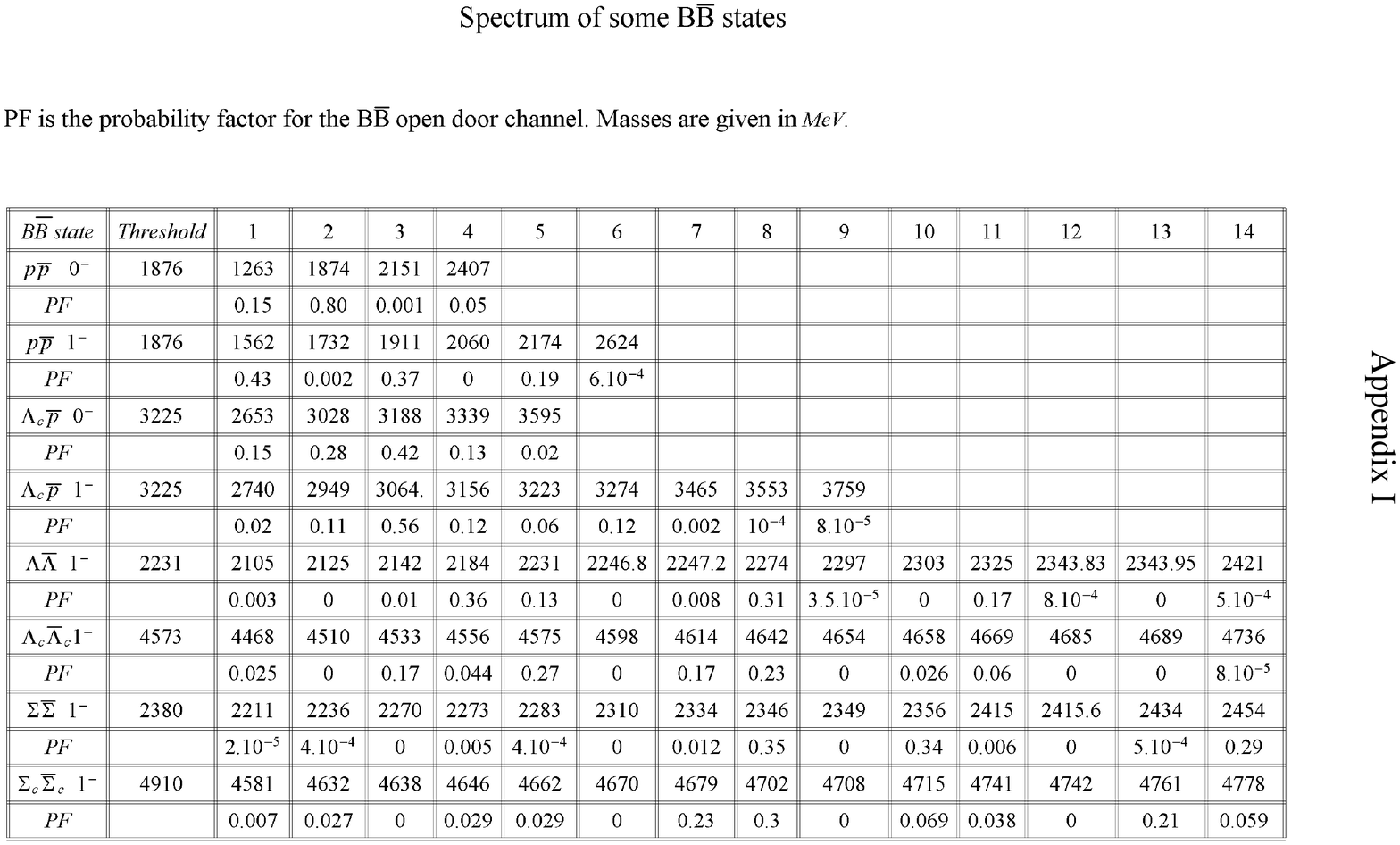}
\end{figure*}
\end{center}
\end{widetext}
\section{Crossing Matrices}
\begin{center}
{\bf Spin 0}
\end{center}
\begin{center}
\begin{equation} \label{cros01}
R_{\phi \rightarrow \alpha}=\left(
\begin{array}[c]{llll}
\frac{1}{\sqrt{2}} & \frac{1}{2\sqrt{3}} & \frac{1}{\sqrt{6}} & \frac{1}{2}\\
-\frac{1}{\sqrt{6}} & \frac{1}{2} & \frac{1}{\sqrt{2}} & -\frac{1}{2\sqrt{3}}\\
\frac{1}{2} & -\frac{1}{\sqrt{6}} & \frac{1}{2\sqrt{3}} & -\frac{1}{\sqrt{2}}\\
-\frac{1}{2\sqrt{3}} & -\frac{1}{\sqrt{2}} & \frac{1}{2} & \frac{1}{\sqrt{6}}
\end{array}
\right)
\end{equation}
\end{center}
\begin{center}
\begin{equation} \label{cros02}
R_{\phi \rightarrow \epsilon}=\left(
\begin{array}
[c]{llll}
\frac{1}{\sqrt{2}} & \frac{1}{2\sqrt{3}} & -\frac{1}{\sqrt{6}} & -\frac{1}{2}\\
-\frac{1}{\sqrt{6}} & \frac{1}{2} & -\frac{1}{\sqrt{2}} & \frac{1}{2\sqrt{3}}\\
\frac{1}{2} & -\frac{1}{\sqrt{6}} & -\frac{1}{2\sqrt{3}} & \frac{1}{\sqrt{2}}\\
-\frac{1}{2\sqrt{3}} & -\frac{1}{\sqrt{2}} & -\frac{1}{2} & -\frac{1}{\sqrt{6}}
\end{array}
\right)
\end{equation}
\end{center}
\begin{center}
{\bf Spin 1}
\end{center}
\begin{center}
\begin{equation} \label{cros11}
R_{\psi \rightarrow \beta}=\left(
\begin{array}[c]{llllll}
\frac{1}{\sqrt{3}} & \frac{1}{\sqrt{6}} & \frac{1}{2\sqrt{3}} & -\frac
{1}{\sqrt{6}} & -\frac{1}{2\sqrt{3}} & \frac{1}{\sqrt{6}}\\
\frac{1}{\sqrt{3}} & \frac{1}{\sqrt{6}} & -\frac{1}{2\sqrt{3}} & \frac
{1}{\sqrt{6}} & \frac{1}{2\sqrt{3}} & -\frac{1}{\sqrt{6}}\\
0 & 0 & \frac{1}{\sqrt{6}} & \frac{1}{\sqrt{3}} & \frac{1}{\sqrt{6}} &
\frac{1}{\sqrt{3}}\\
\frac{1}{\sqrt{6}} & -\frac{1}{\sqrt{3}} & -\frac{1}{\sqrt{6}} & -\frac
{1}{2\sqrt{3}} & \frac{1}{\sqrt{6}} & \frac{1}{2\sqrt{3}}\\
\frac{1}{\sqrt{6}} & -\frac{1}{\sqrt{3}} & \frac{1}{\sqrt{6}} & \frac
{1}{2\sqrt{3}} & -\frac{1}{\sqrt{6}} & -\frac{1}{2\sqrt{3}}\\
0 & 0 & -\frac{1}{\sqrt{3}} & \frac{1}{\sqrt{6}} & -\frac{1}{\sqrt{3}} &
\frac{1}{\sqrt{6}}
\end{array}
\right)
\end{equation}
\end{center}
\begin{center}
\begin{equation} \label{cros12}
R_{\psi \rightarrow \chi}=\left(
\begin{array}[c]{llllll}
\frac{1}{\sqrt{3}} & -\frac{1}{\sqrt{6}} & \frac{1}{2\sqrt{3}} & -\frac
{1}{\sqrt{6}} & \frac{1}{2\sqrt{3}} & -\frac{1}{\sqrt{6}}\\
\frac{1}{\sqrt{3}} & -\frac{1}{\sqrt{6}} & -\frac{1}{2\sqrt{3}} & \frac
{1}{\sqrt{6}} & -\frac{1}{2\sqrt{3}} & \frac{1}{\sqrt{6}}\\
0 & 0 & -\frac{1}{\sqrt{6}} & -\frac{1}{\sqrt{3}} & \frac{1}{\sqrt{6}} &
\frac{1}{\sqrt{3}}\\
\frac{1}{\sqrt{6}} & \frac{1}{\sqrt{3}} & -\frac{1}{\sqrt{6}} & -\frac
{1}{2\sqrt{3}} & -\frac{1}{\sqrt{6}} & -\frac{1}{2\sqrt{3}}\\
\frac{1}{\sqrt{6}} & \frac{1}{\sqrt{3}} & \frac{1}{\sqrt{6}} & \frac{1}
{2\sqrt{3}} & \frac{1}{\sqrt{6}} & \frac{1}{2\sqrt{3}}\\
0 & 0 & \frac{1}{\sqrt{3}} & -\frac{1}{\sqrt{6}} & -\frac{1}{\sqrt{3}} &
\frac{1}{\sqrt{6}}
\end{array}
\right)
\end{equation}
\end{center}
\begin{center}
{\bf Spin 2}
\end{center}
\begin{table}[h]
\begin{tabular}{lr}
$
R_{\xi \rightarrow \gamma}=
\frac{1}{\sqrt{3}} \left(
\begin{array}[c]{cc}
\sqrt{2} & 1\\
1 & -\sqrt{2}
\end{array}
\right)
$&$
R_{\xi \rightarrow \delta}=\frac{1}{\sqrt{3}}\left(
\begin{array}
[c]{cc}%
\sqrt{2} & -1\\
-1 & -\sqrt{2}
\end{array}
\right)
$
\\
\end{tabular}
\end{table}
\begin{equation} \label{cros2}
\end{equation}

\section{Chromo-magnetic operator for $qqq\overline{q}\overline{q}\overline{q}$ states}

We have computed the matrices of chromo-magnetism by inserting the operator Eq.~\ref{eq:HCM}
between the states at Eqs.~[\ref{eqs0:6},\ref{eqs1:6}], they are given below, where $\mathbf{A_{0}}$
is for $0^- $ and $\mathbf{A_{1}}$ for $1^-$.
\begin{center}
\begin{equation} \label{nps0}
{\bf A_{0}}=
\left(
\begin{array}{cccc}
    -2    &      -\sqrt{2}     &     -1              &       0     \\
-\sqrt{2} &         -1         & -\frac{3}{\sqrt{2}} & -\sqrt{5}   \\
    -1    & -\frac{3}{\sqrt{2}} &     -2              & -\sqrt{\frac{5}{2}}\\
    0     &      -\sqrt{5}     & -\sqrt{5/2}         & 0
\end{array}
\right)
\end{equation}
\end{center}
\begin{center}
\begin{equation} \label{nps1}
{\bf A_{1}}=
\left(
\begin{array}{llllll}
2 & -\frac{\sqrt{2}}{3} & \frac{\sqrt{5}}{3} & 0 & \frac{2\sqrt{2}}{3} &
-\frac{2\sqrt{2}}{3}\\
-\frac{\sqrt{2}}{3} & \frac{1}{3} & \sqrt{\frac{5}{2}} & -\frac{\sqrt{5}}{3} &
-\frac{1}{3} & \frac{1}{3}\\
\frac{\sqrt{5}}{3} & \sqrt{\frac{5}{2}} & \frac{4}{3} & \frac{5}{3\sqrt{2}} &
\frac{\sqrt{\frac{5}{2}}}{3} & -\frac{\sqrt{\frac{5}{2}}}{3}\\
0 & -\frac{\sqrt{5}}{3} & \frac{5}{3\sqrt{2}} & -\frac{4}{3} & -\frac
{2\sqrt{5}}{3} & \frac{2\sqrt{5}}{3}\\
\frac{2\sqrt{2}}{3} & -\frac{1}{3} & \frac{\sqrt{\frac{5}{2}}}{3} &
-\frac{2\sqrt{5}}{3} & \frac{5}{6} & -\frac{1}{2}\\
-\frac{2\sqrt{2}}{3} & \frac{1}{3} & -\frac{\sqrt{\frac{5}{2}}}{3} &
\frac{2\sqrt{5}}{3} & -\frac{1}{2} & \frac{5}{6}
\end{array}
\right)
\end{equation}
\end{center}
\begin{figure*}[h]
\includegraphics[bb=0 0 842 596,scale=.8,angle=90]{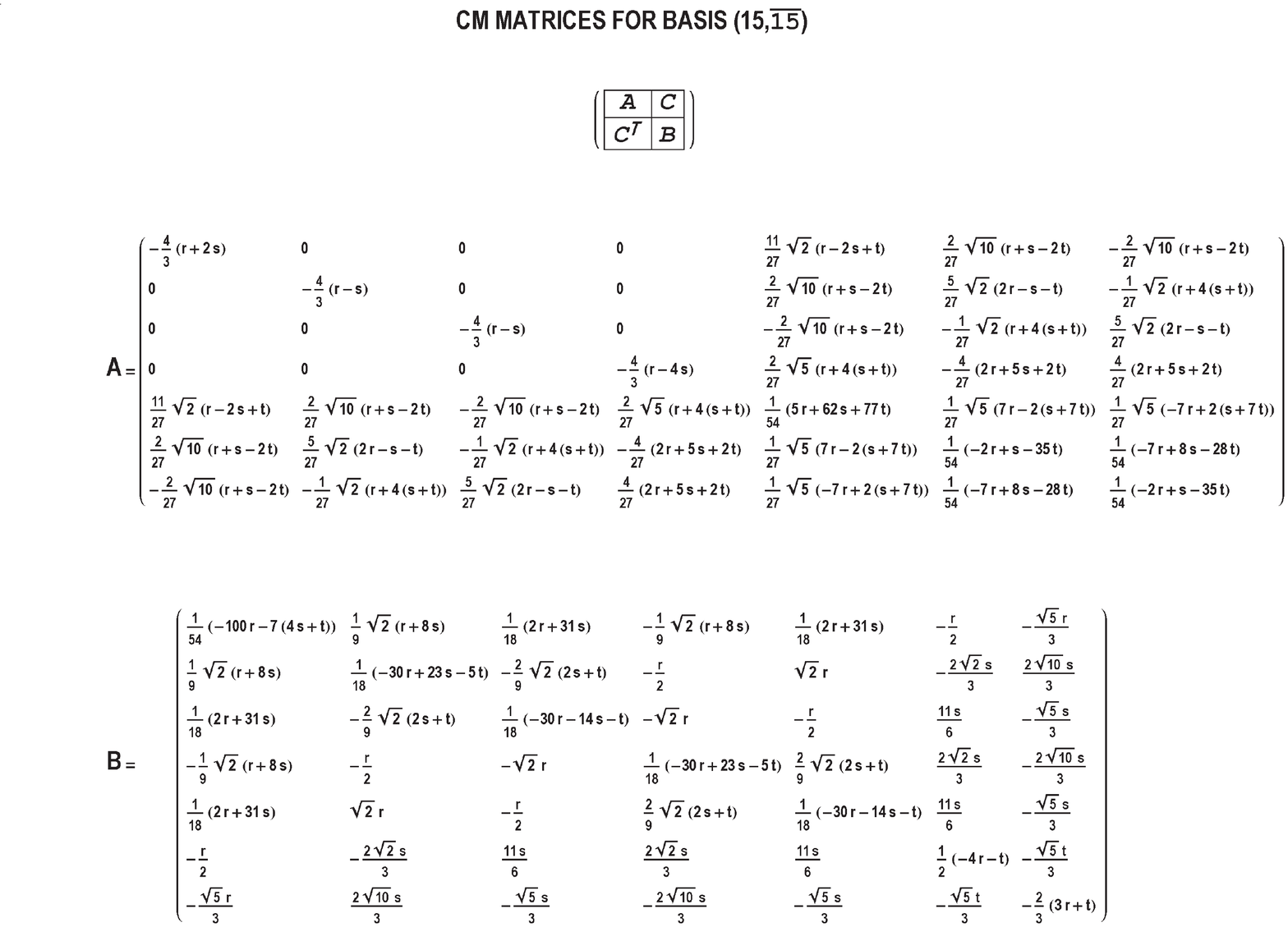}
\end{figure*}
\begin{figure*}[h]
\includegraphics[bb=0 0 842 596,scale=.8,angle=90]{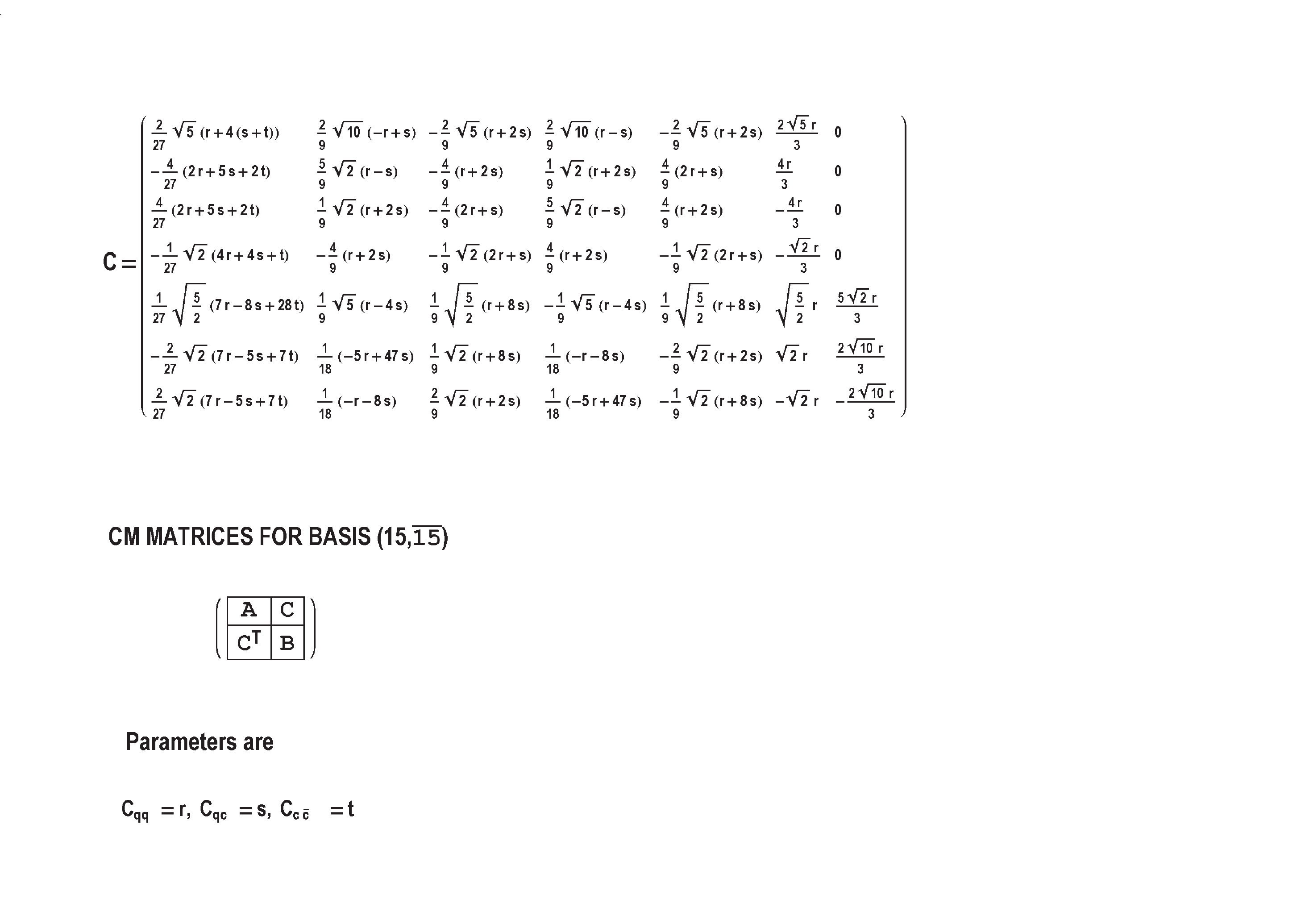}
\end{figure*}
\begin{figure*}[h]
\includegraphics[bb=0 0 842 596,scale=.8,angle=90]{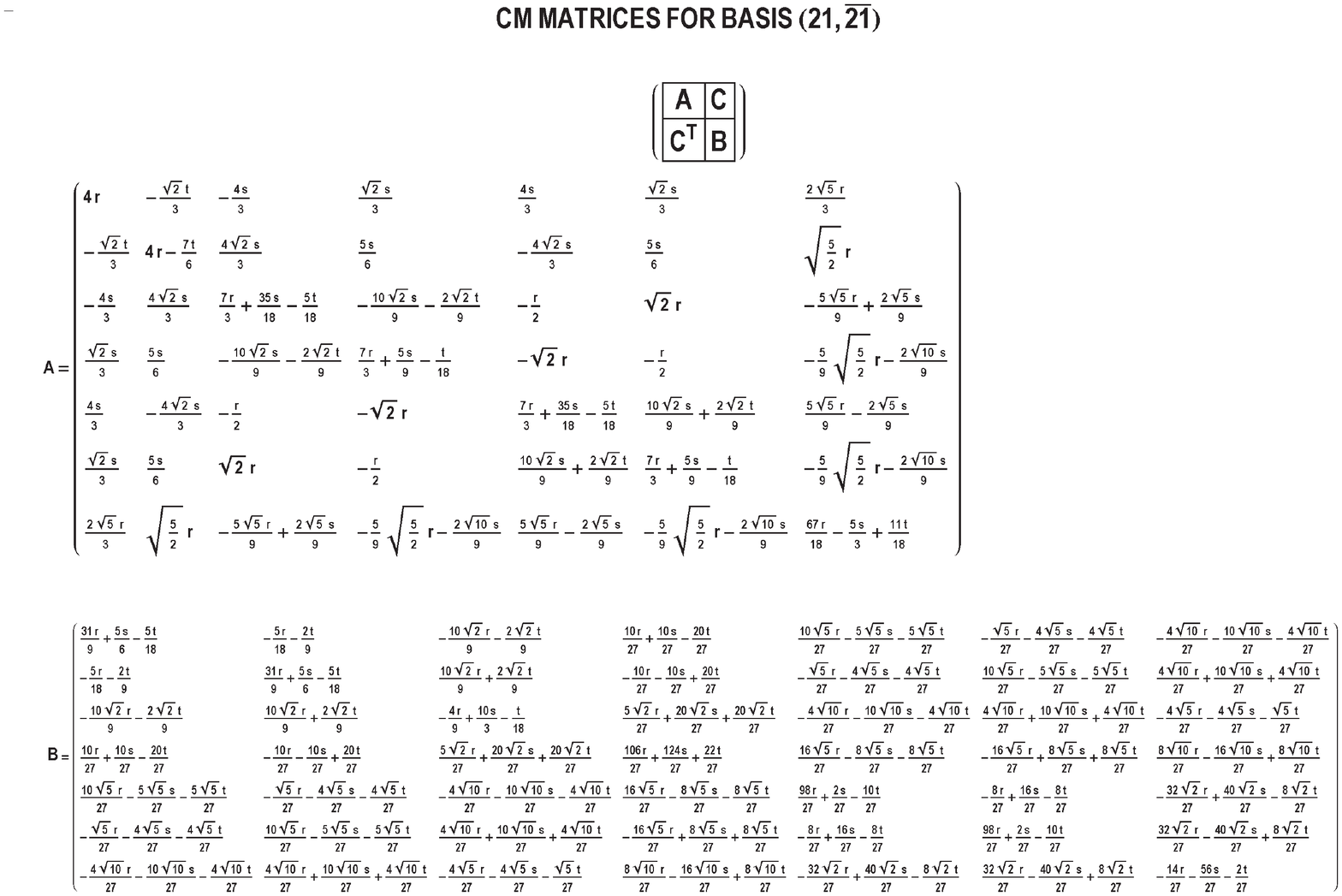}
\end{figure*}
\begin{figure*}[h]
\includegraphics[bb=0 0 842 596,scale=.8,angle=90]{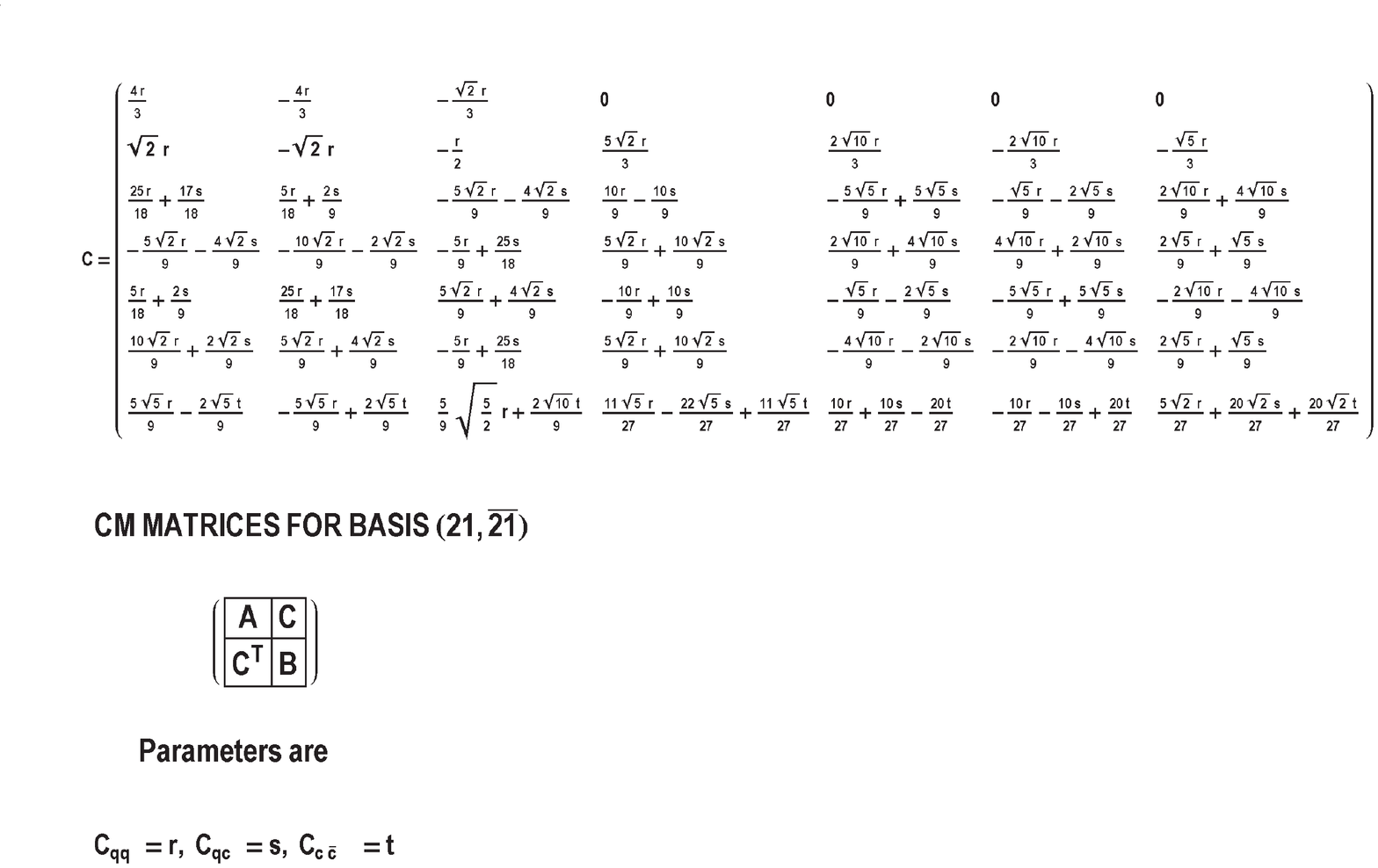}
\end{figure*}
\begin{figure*}[h]
\includegraphics[bb=0 0 842 596,scale=.7,angle=90]{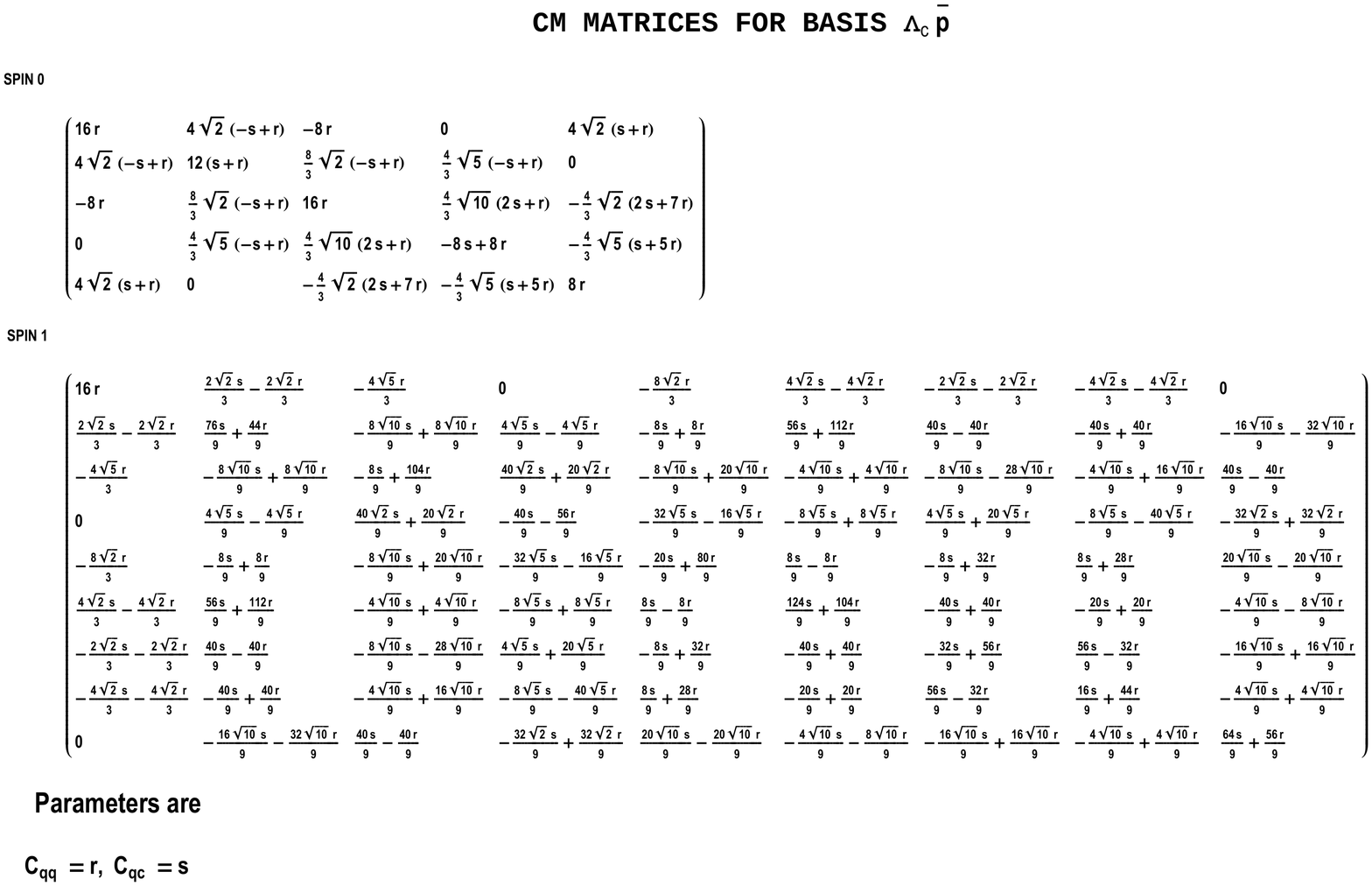}
\end{figure*}

\begin{thebibliography}{99}

\bibitem{JafPHEN1}
R.L. Jaffe, \textit{Phys. Rev.} \textbf{D15} (1977) 267

\bibitem{J}
 R. L. Jaffe, \textit{Phys. Rev.} \textbf{D17} (1978) 1444.
 
\bibitem{JafPHEN2}
R.L. Jaffe, \textit{Phys. Rev.} \textbf{D15} (1977) 281

\bibitem{HS}
 H. H\"{o}gaasen and P. Sorba, \textit{Nucl. Phys.} \textbf{B145} (1978) 119;

M. De Crombrugghe, H. H\"{o}gaasen and P. Sorba, \textit{Nucl. Phys.} 
\textbf{B156} (1979) 347.

\bibitem{JW}  R. L. Jaffe and F. Wilczek, \textit{Phys. Rev. Lett.}
\textbf{91} (2003) 232003; 0307341

\bibitem{MPPR} 
 L.Maiani, F. Piccinini, A. D. Polosa and V.Riquer \textit{Phys.
Rev. Lett.} \textbf{92} (2004) 042003;

\bibitem{KLOE} 
KLOE Collaboration, F. Ambrosino et al
Study of the $a_0(980)$ meson via the radiative decay $\phi->\eta \pi^0$ gamma with the KLOE detector
arXiv:0904.2539v2 [hep-ex]

\bibitem{BigiMaiani}
I.Bigi, L.Maiani, F. Piccinini, A. D. Polosa and V.Riquer,
\textit{Phys. Rev.} \textbf{D72} (2005) 114016,

\bibitem{DGG}  
A. De R\'{u}jula, H. Georgi and S.L. Glashow, \textit{Phys.
Rev.} \textbf{D12} (1975) 147.

\bibitem{BHRS}  
F. Buccella , H. H\"{o}gaasen, J. M. Richard and P. Sorba, 
\textit{Eur. Phys. J.} \textbf{C49} (2007) 743

\bibitem{Bai06} 
BES Collaboration: J.Z.Bai, Y.Ban, J.G.Bian 
\textit{Phys. Rev. Lett.} \textbf{91} (2003) 022001
arXiv:hep-ex/0303006 

\bibitem{Gaby}
 Gabyshev et al., \textit{Phys. Rev. Lett.}
\textbf{97} (2006) 242001 
arXiv:hep-ex/0409005v4

\bibitem{Y4660}
G. Pakhlova et al. (Belle Collaboration), \textit{Phys. Rev. Lett.} \textbf{101}
(2008) 172001

\bibitem {JafPR}
R.L. Jaffe, \textit{Phys. Rept.} \textbf{409} (2005) 1;

\textit{Nucl. Phys. Proc. Suppl.} \textbf{142} (2005) 343

\bibitem{Sch08}
Amir H. Fariborz, Renata Jora, Joseph Schechter,
\textit{Phys. Rev.} \textbf{D77} (2008) 094004
arXiv:0801.2552v2 [hep-ph]

\bibitem{form}
J.~A.~M.~Vermaseren, arXiv:math-ph/0010025.

\bibitem{ABFRT}
M.Abud,F.~Buccella, D.~Falcone, G. Ricciardi and
F.~Tramontano, \textit{Adv. Studies Theor. Phys.} \textbf{2} (2008) 929.

\bibitem{ETAb} 
Aubert B et al., \textit{Phys. Rev. Lett.} \textbf{101} (2008) 071801,
see also

Joerg Marks, for the BABAR Collaboration,
Bottomonium Results from BABAR and BELLE
arXiv:0906.0725v1 [hep-ex]
\bibitem{QuiggRossner}
C. Quigg and J.L. Rosner,
\textit{Phys. Rept.} \textbf{56} (1979) 167

\bibitem{Gatto}
R.Casalbuoni, A. Deandrea, N.Di Bartolomeo, F.Feruglio, R.Gatto, G. Nardulli,
\textit{Phys. Rept.} \textbf{281} (1997) 145
arXiv:hep-ph/9605342 

\bibitem{Albrecht1991F}
Albrecht et al,
\textit{Z. Phys.} \textbf{C50} (1991) 1
 
\bibitem{Etkin:1987rj}
Etkin, A et al,
\textit{Phys. Lett.} \textbf{B201} (1988) 568

\bibitem{PDG} 
C. Amsler et al. (Particle Data Group), \textit{Phys. Lett.} \textbf{B667} (2008) 1
and 2009 partial update for the 2010 edition. 

\bibitem{Gasp} 
M. Gaspero,\textit{Nucl. Phys.} \textbf{A562} (1993) 407.

\bibitem{Hooft}
 G. 't Hooft,  G. Isidori,  L. Maiani, A. D. Polosa  and V.Riquer 
\textit{Phys. Lett.} \textbf{B662} (2008) 424.

\bibitem{Y(4140)}
T. Aaltonen et al. (CDF Collab.), Phys.Rev.Lett. \textbf{102},  242002 (2009).

\bibitem{Stancu}
Fl.Stancu,
arXiv0906.2485 [hep-ph]

\bibitem{X(4350)}
C. P. Shen et al, for the Belle Collaboration,
arXiv:0912.2383v1 [hep-ex]

\bibitem{a01330}
S. Uehara, et al., for the Belle Collaboration 
\textit{Phys. Rev.} \textbf{D80} (2009) 032001
arXiv:0906.1464 

\bibitem{BESgammaX}
BES Collaboration, M. Ablikim et al,
\textit{Phys. Rev. Lett.} \textbf{95} (2005) 262001
arXiv:hep-ex/0508025

\bibitem{Eva}
C.Evangelista et al. \textit{Nucl. Phys.} \textbf{B153} (1979) 253;

\bibitem{BABARY}
B. Aubert et al., BaBar Collaboration, \textit{Phys. Rev.} \textbf{D74} (2006) 091103

\bibitem{BES4} 
BES Collaboration, M. Ablikim et al, \textit{Phys. Rev. Lett.} \textbf{100} (2008) 102003
arXiv:0712.1143v5 [hep-ex]

\bibitem{DETROIT}
A. Zupanc (for the Belle Collaboration),
Hadron Spectroscopy Results from Belle,
To be published in the proceedings of DPF-2009, Detroit, MI, July 2009, eConf C090726
arXiv:0910.3404v1 [hep-ex]

\bibitem{PSIPIPI} 
X. L. Wang, et al, Belle Collaboration,
\textit{Phys. Rev. Lett.} \textbf{99} (2007) 142002
arXiv:0707.3699v2 [hep-ex]

\bibitem{Choi}
S.K. Choi, Talk given at the 34th International Conference on High Energy Physics,
Philadelphia, 2008 arXiv:0810.3546v1 [hep-ex]

\bibitem{QED}
G. Cotugno, R. Faccini, A. D. Polosa, C. Sabelli 
arXiv:0911.2178
\end{thebibliography}
\end{document}